\documentclass[journal]{IEEEtran}
\usepackage{epsfig,color,amsmath,cite}
\usepackage{amsthm}
\usepackage{wrapfig}
\usepackage{bm}
\usepackage{epstopdf}
\usepackage{amssymb}
\usepackage{url}
\usepackage{enumitem}
\usepackage{multirow}
\usepackage{graphicx}
\usepackage{color}
\usepackage{amsfonts}
\usepackage{float}
\usepackage{amsmath}
\usepackage{amsthm}
\usepackage{epstopdf}
\usepackage{changepage}
\usepackage{cases}
\usepackage{booktabs}
\usepackage{footnote}
\makesavenoteenv{tabular}
\usepackage{threeparttable}
\usepackage[linesnumbered,lined,boxed,commentsnumbered,ruled,longend]{algorithm2e}


\setlist[itemize]{leftmargin=*}

\DeclareMathOperator*{\minimize}{minimize}

\DeclareMathOperator{\subjectto}{subject\ to}

\makeatother

\newtheorem{theorem}{Theorem}
\newtheorem{mydef}{Definition}

\newcommand{\mat}[1]{\boldsymbol{#1}}

\newcommand{\bmat}[1]{\begin{bmatrix} #1 \end{bmatrix}}

\providecommand{\mA}{\ensuremath{\mat{A}}}

\providecommand{\mC}{\ensuremath{\mat{C}}}

\providecommand{\mI}{\ensuremath{\mat{I}}}

\providecommand{\mL}{\ensuremath{\mat{L}}}

\providecommand{\mO}{\ensuremath{\mat{O}}}
\providecommand{\mP}{\ensuremath{\mat{P}}}
\providecommand{\mQ}{\ensuremath{\mat{Q}}}

\providecommand{\mY}{\ensuremath{\mat{Y}}}
\providecommand{\mZ}{\ensuremath{\mat{Z}}}

\newcommand{\m}{\boldsymbol}
\allowdisplaybreaks[4]
\usepackage[colorlinks = true,
linkcolor = red,
urlcolor  = blue,
citecolor = blue,
anchorcolor = blue]{hyperref}

\newcommand{\mc}[1]{\mathcal{#1}}

\usepackage[noabbrev]{cleveref}

\usepackage{mathtools}

\DeclarePairedDelimiter\abs{\lvert}{\rvert}%
\DeclarePairedDelimiter\norm{\lVert}{\rVert}%

\makeatletter
\let\oldabs\abs
\def\abs{\@ifstar{\oldabs}{\oldabs*}}
\let\oldnorm\norm
\def\norm{\@ifstar{\oldnorm}{\oldnorm*}}
\makeatother

\DeclareMathAlphabet\mathbfcal{OMS}{cmsy}{b}{n}

\usepackage{stackengine}
\newcommand\barbelow[1]{\stackunder[1.2pt]{$#1$}{\rule{.8ex}{.075ex}}}

\usepackage{graphicx}
\usepackage{float}
\usepackage[caption = false]{subfig}

\usepackage[english]{babel}
\usepackage[utf8]{inputenc}
\usepackage[super]{nth}

\usepackage[dvipsnames]{xcolor}

\usepackage{framed}

\usepackage{booktabs}
\begin{document}
	
	\newdimen\origiwspc%
	\newdimen\origiwstr%
	\origiwspc=\fontdimen2\font
	\origiwstr=\fontdimen3\font
	
	\fontdimen2\font=0.63ex 
	
\title{\textcolor{black}{Robust Dynamic State Estimation of Synchronous Machines with Asymptotic State Estimation Error Performance Guarantees}}
	\author{\hspace{-0.0cm}Sebastian A. Nugroho,~\IEEEmembership{Student Member,~IEEE,} Ahmad F. Taha,~\IEEEmembership{Member,~IEEE,} and
		 Junjian~Qi,~\IEEEmembership{Senior Member,~IEEE}
		\thanks{S. A. Nugroho and A. F. Taha are with the Department of Electrical and Computer Engineering, the University of Texas at San Antonio, San Antonio, TX 78249 (e-mails: sebastian.nugroho@my.utsa.edu, ahmad.taha@utsa.edu).
		J.~Qi is with the Department of Electrical and Computer Engineering, University of Central Florida, Orlando, FL 32816 USA (e-mail: Junjian.Qi@ucf.edu). S. Nugroho and A. F. Taha's work was supported by the National Science Foundation under Grants 1728629 and 1917164. J. Qi's work was supported by Cyber Florida under Collaborative Seed Award 3910-1009-00.	}
	}
	\maketitle

\begin{abstract}
\textcolor{black}{A robust observer for performing power system dynamic state estimation (DSE) of a synchronous generator is proposed. The observer is developed using the concept of $\mathcal{L}_{\infty}$ stability for uncertain, nonlinear dynamic generator models. We use this concept to \textit{(i)} design a simple, scalable, and robust dynamic state estimator and \textit{(ii)} obtain a performance guarantee on the state estimation error norm relative to the magnitude of uncertainty from unknown generator inputs, and process and measurement noises. Theoretical methods to obtain upper and lower bounds on the estimation error are also provided. Numerical tests validate the performance of the $\mathcal{L}_{\infty}$-based estimator in performing DSE under various scenarios. The case studies reveal that the derived theoretical bounds are valid for a variety of case studies and operating conditions, while yielding better performance than existing power system DSE methods.}
\end{abstract}


\section{Introduction}
	
\IEEEPARstart{P}{ower} system monitoring and operation rely on static state estimation (SSE) \cite{se3,se2,MNMR}, which assumes the system is operating in quasi-steady state and estimates its static states, the voltage magnitude and phase angles, using SCADA and/or phasor measurement unit (PMU) data \cite{gol2015hybrid,fernandes2017application}. However, power systems do not operate in steady state mainly due to load variations and potential large disturbances. They are becoming even more dynamic with the increasing integration of utility-scale renewable generation to transmission grids and a large number of distributed energy resources to distribution grids. 
	
Therefore, SSE may not be sufficient in rapidly and accurately capturing the dynamic states of the system for desirable situational awareness. By contrast, with the wide-spread deployments of PMUs, the dynamic state estimation (DSE) could effectively estimate the dynamic states (i.e., the internal states of generators) based on the available model of the dynamic components and PMU measurements that have high time-synchronization accuracy and high sampling rates. 

To enable real-time system monitoring, protection, and control, DSE has to be time-efficient enough to allow for online application. DSE should also be robust against model uncertainties (i.e. un-modeled dynamics, unknown inputs, inaccurate parameters), measurement gross errors (i.e. bad/low-quality/missing data or manipulated measurements by cyber attacks), and non-Gaussian measurement noises, considering that extensive results using field PMU data from WECC system have revealed that the Gaussian assumption is questionable \cite{error}.
	
DSE has been implemented by several stochastic estimators. For example, \textit{extended Kalman filter} (EKF) is applied to perform DSE \cite{ekf1,ghahremani2011dynamic}, which works  only  in a mild nonlinear environment and when Jacobian matrix exists.
As a derivative-free alternative, \textit{unscented Kalman filter} (UKF) does not require linearization or calculation of Jacobian matrices \cite{valverde2011unscented,pwukf1,pwukf2,qi,sun2016power,qi2015dynamic}. 
Using spherical-radial cubature rule, Arasaratnam \textit{et al.} propose \textit{cubature Kalman filter} (CKF) \cite{CKF}, which has been shown to have improved performance compared with EKF and UKF \cite{access_18}. 
Besides, extended particle filter \cite{zhou,cui2015particle} and ensemble Kalman filter \cite{zhou2015dynamic} are also applied to perform DSE. 

In order to deal with model uncertainty, gross errors, or non-Gaussian noises, robust DSE has been proposed. 
Generalized maximum-likelihood-type estimate \cite{gandhi2010robust}, two-stage Kalman filter \cite{zhang2014two}, iterated EKF \cite{zhao1}, $H_\infty$ EKF \cite{zhao2}, robust UKF \cite{zhao3}, and robust CKF \cite{li_ckf} have been developed. 
\textcolor{black}{In \cite{zhao2019constrained}, a constrained robust DSE method is proposed that deals with both equality and inequality constraints for state variables and parameters.} Unlike most papers such as  \cite{ghahremani2011online,zhou,zhou2015dynamic} that assume all inputs to a synchronous machine are measurable, in \cite{ekf2,ghahremani2014synchrophasor} an EKF-unknown input (EKF-UI) method is implemented for  
one synchronous machine, assuming some or all inputs are unknown, which is more realistic since some inputs such as the mechanical torque can be hard to measure \cite{pwukf2,ekf2}. In \cite{anagnostou2018derivative}, a derivative-free Kalman filter considering unknown inputs is proposed. However, in \cite{ekf2,ghahremani2014synchrophasor} it is required that the number of output measurements be greater than the number of unknown inputs.  Also, the above Kalman filter-based methods do not have theoretical guarantee for convergence. In addition to the approaches discussed above, other approaches, mostly based on different variants of Kalman filter have also been proposed. A thorough review of different approaches on DSE and robust DSE can be found in \cite{zhao2019power} and \cite{singh2018dynamic}. 

Apart from stochastic estimators, deterministic observers that do not require noise distribution have also been designed for DSE in power systems. While observers are often used in observer-based control architectures and applications, stochastic estimators are typically used for monitoring. With that in mind, stochastic estimators and dynamic observers both perform the same task: estimating states of a noisy or uncertain dynamical system with limited number of sensors~\cite{taha2015secure}. In \cite{7491374} a sliding-mode observer and an attack detection filter are proposed to estimate the power system's unknown inputs and detect potential cyber attacks. The authors in~\cite{Jin2018} develop a multiplier-based observer design for Lipschitz systems and utilize the designed observer on a multi-machine power system with second-order classical generator model and linear measurement model. 
In \cite{access_18} a thorough experimental comparison between stochastic estimators and deterministic observers is performed under various classes of uncertainty. Recently, an anomaly detection algorithm for detecting changes in power system operational range for synchronous generators is introduced in \cite{Anagnostou2018}, in which the detection algorithm uses the predicted states obtained from the proposed time-varying observer.  

In this paper we propose a new observer design for performing decentralized DSE on synchronous generators. The observer operates based on the nonlinear dynamics of the generator, which is subject to process and measurement noise. Specifically, we also assume that some of the generator inputs such as mechanical torque and internal field voltage are unknown to the observer. The contributions of the paper are given as follows.
\begin{itemize}
	{\color{black}
    \item Based on the Lipschitz property of the generator's nonlinear model and PMU measurements \cite{Nugroho2019}, we propose a new robust observer framework using the concept of $\mc{L}_{\infty}$ stability that provides a performance guarantee for the state estimation error norm against worst case disturbance (due to uncertainty in generator inputs and noise). This performance can be assessed from a constant called the \textit{performance index or level}. Albeit the observer design is originally posed as a nonconvex semidefinite program (SDP), we show that the observer gain matrix can be computed via solving a convex optimization problem. Moreover, as the observer gain remains constant, the observer is very efficient for real-time implementation.
    
    \item We introduce a scalable computational method to obtain lower and upper bounds for the optimal performance index. The lower bound is derived based on an SDP relaxation, and the upper bound is derived using simple methods. The bounds are useful in assessing and quantifying the optimality of the performance index and hence the state estimation quality.

    \item The proposed $\mathcal{L}_{\infty}$ observer for power system DSE is tested under various case studies which include: \textit{(a)} numerous noise distributions and magnitudes, \textit{(b)} high-order generator dynamic models, and \textit{(c)} comparison with some Kalman filter-based state estimators.   
}
\end{itemize}

The case studies reveal that the derived theoretical bounds are valid for a variety of case studies and operating conditions, while also yielding better performance than mainstream power system DSE methods under the aforementioned scenarios and case studies. The remainder of the paper is organized as follows. Section~\ref{sec:model} presents the nonlinear process-measurement generator model with a PMU installed at the terminal bus of the generator. Section~\ref{sec:observer-design} focuses on the design of  robust $\mathcal{L}_{\infty}$ observer. In Section \ref{sec:bounding_SDP}, we propose methods to pose the $\mathcal{L}_{\infty}$ observer design as convex SDP and  compute lower and upper bounds for the optimal performance index. Section~\ref{sec:simulations} presents a comprehensive numerical experiments and Section \ref{conclusion} concludes the paper.

\section{Generator Dynamic Model Under Uncertainty}  \label{sec:model}

For DSE, both multi-machine model \cite{ekf1,pwukf1,zhang2014two,qi,qi2015dynamic,access_18,zhao1,zhao2,zhao3} and single-machine model \cite{ghahremani2011online,ekf2,ghahremani2014synchrophasor,zhou,zhou2015dynamic,rouhani2017observability} have been used. Compared with the centralized estimation approach based on multi-machine model, using the one-machine model and modeling the rest of the system as inputs can enable decentralized estimation, reducing computational complexity and avoiding to send all data through dedicated communication network to control center. 
{\color{black}
With that in mind, in this section we focus on modeling and understanding the nonlinearities of a single synchronous machine, whose transient model can be described by the following \nth{4}-order differential equations in local $\mathrm{d}$-$\mathrm{q}$ reference frame~\cite{sauer1998power,kundur1994power}}.
\begin{subnumcases}{\label{gen model}}
	\dot{\delta}=\omega-\omega_0 \\
	\dot{\omega}=\frac{\omega_0}{2H_i}\Big(T_{\textrm{m}}-T_{\textrm{e}}-\frac{K_{\textrm{D}}}{\omega_0}(\omega-\omega_0)\Big) \\
	\dot{e}'_{\textrm{q}}=\frac{1}{T'_{\textrm{d0}}}\Big(E_{\textrm{fd}}-e'_{\textrm{q}}-(x_{\textrm{d}}-x'_{\textrm{d}})\,i_{\textrm{d}}\Big) \\
	\dot{e}'_{\textrm{d}}=\frac{1}{T'_{\textrm{q0}}}\Big(-e'_{\textrm{d}}+(x_{\textrm{q}}-x'_{\textrm{q}})\,i_{\textrm{q}}\Big),
	\end{subnumcases}
	where 
	$\delta(t):=\delta$ is the rotor angle,
	$\omega(t):=\omega$ is the rotor speed in rad/s, and $e'_{\mathrm{q}}(t):=e'_{\mathrm{q}}$ and $e'_{\mathrm{d}}(t):=e'_{\mathrm{d}}$ are the transient voltage along $\mathrm{q}$ and $\mathrm{d}$ axes; 
	$i_{\mathrm{q}}(t):=i_{\mathrm{q}}$ and $i_{\mathrm{d}}(t):=i_{\mathrm{d}}$ are stator currents at $\mathrm{q}$ and $\mathrm{d}$ axes;
	$T_{\mathrm{m}}(t):=T_{\mathrm{m}}$ is the mechanical torque, $T_{\mathrm{e}}(t):=T_{\mathrm{e}}$ is the electric air-gap torque, 
	and $E_{\mathrm{fd}}(t):=E_{\mathrm{fd}}$ is the internal field voltage; $\omega_0=2\pi 60 \mathrm{rad}/\mathrm{sec}$ is the rated value of angular frequency, $H$ is the inertia constant, and $K_{\mathrm{D}}$ is the damping factor; 
	$T'_{\mathrm{q0}}$ and $T'_{\mathrm{d0}}$ are the open-circuit time constants for $\mathrm{q}$ and $\mathrm{d}$ axes; $x_{\mathrm{q}}$ and $x_{\mathrm{d}}$ are the synchronous reactance and $x'_{\mathrm{q}}$ and $x'_{\mathrm{d}}$ are the transient reactance respectively at the $\mathrm{q}$ and $\mathrm{d}$ axes. \textcolor{black}{Case studies also consider higher-order generator models.}
	
	We assume that a PMU is installed at the terminal bus of the generator. The mechanical torque $T_{\mathrm{m}}$ and internal field voltage $E_{\mathrm{fd}}$ are considered as \textit{unknown inputs}, which values are assumed to be unknown.
	Additionally, we take the current phasor $I_{t}=i_{\mathrm{R}} + j i_{\mathrm{I}}$ measured by PMU as inputs which can help decouple the generator from the rest of the network \cite{zhou}. The voltage phasor $E_{t}=e_{\mathrm{R}} + j e_{\mathrm{I}}$ can also be measured by PMU and is considered as output. 
	The dynamic model (\ref{gen model}) can be expressed in a general state space form where the state, input, unknown input, and output vectors are respectively defined as $\m x = \bmat{\delta \quad \omega \quad e'_{\mathrm{q}} \quad e'_{\mathrm{d}}}^\top,$ $\m u = \bmat{ i_{\mathrm{R}} \quad i_{\mathrm{I}}}^\top,$ ${\m q} =  \bmat{T_{\mathrm{m}} \quad E_{\mathrm{fd}} }^\top,$ and $\m y = \bmat{e_{\mathrm{R}} \quad e_{\mathrm{I}}}^\top$. The $i_{\textrm{q}}$, $i_{\textrm{d}}$, and $T_{\textrm{e}}$ in (\ref{gen model}) are functions of $\m{x}$ and $\m{u}$ given as follows
	\vspace{-0.1cm}
	\begin{align}
	\vspace{-0.1cm}
	&i_{\textrm{q}}=u_2 \sin x_1 + u_1 \cos x_1,\quad 
	i_{\textrm{d}}=u_1 \sin x_1-u_2 \cos x_1\notag \\
	&e_{\textrm{q}}= x_3-\frac{S_\textrm{B}}{S_{\textrm{N}}}x'_{\textrm{d}}i_{\textrm{d}} , \quad
	e_{\textrm{d}}= x_4 + \frac{S_\textrm{B}}{S_{\textrm{N}}}x'_{\textrm{q}}i_{\textrm{q}} \notag \\
	&P_{\textrm{e}} = e_{\textrm{q}}i_{\textrm{q}}+e_{\textrm{d}}i_{\textrm{d}} , \quad
	T_{\textrm{e}} = \frac{S_\textrm{B}}{S_{\textrm{N}}} P_{\textrm{e}}, \notag
	\end{align}
	where  $e_{\mathrm{q}}$ and $e_{\mathrm{d}}$ are the terminal voltage at $\mathrm{q}$ and $\mathrm{d}$ axes, and $S_\mathrm{B}$ and $S_{\mathrm{N}}$ are the system base MVA and the base MVA for this generator.
	The PMU outputs $e_{\textrm{R}}$ and $e_{\textrm{I}}$ can be written as functions of $\m{x}$ and $\m{u}$ as follows
	\vspace{-0.1cm}
	\begin{align}
		&y_1= e_{\textrm{d}}\sin\delta+e_{\textrm{q}}\cos\delta , \quad y_2=e_{\textrm{q}}\sin\delta-e_{\textrm{d}}\cos\delta.\label{output equation}
	\end{align}
	To that end, we can rewrite \eqref{gen model} and \eqref{output equation} as 
\begin{subequations}{\label{gen_model1}}
	\begin{align}
	&\dot{x}_1=x_2-\alpha_1 \\
	&\dot{x}_2=\alpha_2q_1-\alpha_5x_2-\alpha_3\left(x_3u_2+x_4u_1\right)\sin x_1\nonumber\\
	&\quad\quad \,+\,\alpha_3\left(x_4u_2-x_3u_1\right)\cos x_1 +\alpha_4u_1u_2\cos 2x_1 \,\quad \nonumber \\
	&\quad\quad \,+\,\tfrac{1}{2}\alpha_4\left(u_2^2-u_1^2\right)\sin 2x_1 + \alpha_6 \\
	&\dot{x}_3=\alpha_7 q_2 - \alpha_7 x_3 - \alpha_8 \left(u_1\sin x_1-u_2\cos x_1\right) \\
	&\dot{x}_4=\alpha_{10} \left(u_1\cos x_1+u_2\sin x_1\right) - \alpha_9 x_4, \\
	&y_1 = x_3\cos x_1 + x_4\sin x_1 +\beta_1u_1\sin 2x_1 \nonumber \\ 
	&\qquad + \beta_1u_2\cos 2x_1 + \beta_2u_2\\
	&y_2 = x_3\sin x_1 - x_4\cos x_1 -\beta_1u_1\cos 2x_1 \nonumber \\ 
	&\qquad - \beta_1u_2\sin 2x_1 - \beta_2u_1,			
	\end{align}
\end{subequations}
where the parameters $\alpha_1,\ldots,\alpha_{10}$, $\beta_1$, and $\beta_2$ are given in Appendix A of \cite{Nugroho2019}. 
	By rearranging the state vector $\m x$ and input vector $\m u$ and separating the linear terms from the nonlinear ones, generator dynamics \eqref{gen_model1} can be expressed as
	\begin{subnumcases} {\label{eq:state_space_gen}}
	\dot{\m{x}}= \mA \m x + \m{f}(\m{x},\m{u}) +\m{{B}_q} {\m q} \label{eq:state_space_gen-a}  \\
	\m{y}=\m{h}(\m{x},\m{u}) + \m{D_u} \m u, \label{eq:state_space_gen-b}
	\end{subnumcases}
where matrices $\m A$, $\m{{B}_q}$, and $\m{D_u}$ are
	\begin{align*}
	\m A = &\bmat{0&1&0&0\\0&-\alpha_5&0&0\\0&0&-\alpha_7&0\\0&0&0&-\alpha_9},\,\;
	\m{{B}_q} = \bmat{0&0\\\alpha_2&0\\0&\alpha_7\\0&0}, \\
	\m{D_u} = &\bmat{0&0&0&\beta_2\\0&0&-\beta_2&0}.\;
	\end{align*}
	Functions $\m f(\cdot)$ and $\m h(\cdot)$ represent the state and output nonlinearities in~\eqref{gen_model1}.
Throughout this paper we assume that the operating region of generator states and inputs are bounded such that $\m x \in \mathbfcal{X}$ and $\m u \in \mathbfcal{U}$ where
\begin{subequations}
	\begin{align}
	\mathbfcal{X} &:= \left[\barbelow{x}_1,\bar{x}_1\right]\times\left[\barbelow{x}_2,\bar{x}_2\right]\times\left[\barbelow{x}_3,\bar{x}_3\right]\times\left[\barbelow{x}_4,\bar{x}_4\right] \\
	\mathbfcal{U} &:= \left[\barbelow{u}_1,\bar{u}_1\right]\times\left[\barbelow{u}_2,\bar{u}_2\right].
	\end{align}
\end{subequations}
\indent Realize that this assumption is practical and holds for most power system models as physical quantities such as angles and frequencies are naturally bounded. 	These upper and lower bounds of the generator states and control inputs, characterized by the sets $\mathbfcal{X}$ and $\mathbfcal{U}$, can be determined from the operator's knowledge of power systems operational range. In addition, $\mathbfcal{X}$ and $\mathbfcal{U}$ may also be determined through performing extensive simulations while applying different faults or contingencies and then finding the operating region of the generator \cite{qi2016nonlinear}. 
	 In the next section, we design a robust observer for DSE which utilizes the fact that $\m f(\cdot)$ and $\m h(\cdot)$ are locally Lipschitz continuous. That is, there exist nonnegative constants $\gamma_f$ and $\gamma_h$ such that
 		\begin{subequations}\label{eq:lipschitz_constants_all}
	 	\begin{align}
	 	\norm{\m f(\m x, \m u)-\m f(\hat{\m x}, \m u)}_2 &\leq \gamma_f \norm{\m x -  \hat{\m x}}_2 \label{eq:lipschitz_constants_all_f}\\
	 	\norm{\m h(\m x, \m u)-\m h(\hat{\m x}, \m u)}_2 &\leq \gamma_h \norm{\m x -  \hat{\m x}}_2. \label{eq:lipschitz_constants_all_h}
	 	\end{align}
	 \end{subequations}
	 The corresponding Lispchitz constants can be computed analytically given that the sets $\mathbfcal{X}$ and $\mathbfcal{U}$ are known. Readers are referred to \cite{Nugroho2019} for a complete derivation of methods to compute $\gamma_f$ and $\gamma_h$ which are a function of state and input bounds as well as the generator state-space matrices.

\vspace*{-0.2cm}
	\section{Robust Observer Design }~\label{sec:observer-design}

	In Section \ref{sec:model} we treat mechanical torque from the turbine $T_{\mathrm{m}}$ and internal field voltage from the exciter $E_{\mathrm{fd}}$ as unknown inputs as these quantities are difficult to measure. Even if $T_{\mathrm{m}}$ and $E_{\mathrm{fd}}$ are measured or estimated, the estimates could still exhibit uncertainty. As a result, here we consider that the robust estimator has access to the entries of $\m{{B}_q}$, but not to the time-varying, uncertain $T_{\mathrm{m}}$ and $E_{\mathrm{fd}}$. This allows unknown input matrices to be constructed as $\m{B_w} := \m{{B}_q}$ and $\m{D_w} := \mO$. For consistency, let $\m{q_w} := \m q$. This allows the generator model \eqref{eq:state_space_gen} to be expressed as
	\begin{subnumcases} {\label{eq:state_space_gen3_lip}}
		\dot{\m{x}}= \mA \m x + \m{f}(\m{x},\m{u}) +\m{{B}_w} \m{q_w}\label{eq:state_space_gen3_lip_a}\\ 
		\m{y}=\m{h}(\m{x},\m{u}) + \m{D_u} \m u  + \m{{D}_w} \m{q_w},\qquad\label{eq:state_space_gen3_lip_b} 
	\end{subnumcases}
where $\m{q_w}(t) \in \mathbb{R}^2$ is the unknown input vector defined above.

\subsection{Model Modification}

The PMU output $\m y(t)$ is nonlinearly related to $\m x(t)$ through $\m h(\m x, \m u)$. To design the robust observer, and inspired by \cite{wang2016}, we introduce the function $\m {h_\mathrm{l}}(\cdot)$ as
\vspace{-0.1cm}
	\begin{align}
	\m{h_\mathrm{l}}(\m x, \m u) &:= -\m{C}\m x + \m h(\m x, \m u). \label{eq:h_l} 
	\end{align}
Since $\m h(\cdot)$ is locally Lipschitz with Lipschitz constant $\gamma_h$, from \eqref{eq:h_l} it follows that $\m{h_\mathrm{l}}(\cdot)$ is also locally Lipschitz with Lipschitz constant $\gamma_l$ where $\gamma_l = \gamma_h + \norm{\mC}_2$. {\color{black} Ideally $\m{C}$ can be any matrix of appropriate dimension, constructed in such a way that the pair $(\mA,\mC)$ is at least detectable. This condition is crucial as it ensures that the internal states of the system can be estimated via observer. See \cite{Anderson1981,chen1999linear} for thorough discussions on detectability/observability.} One straightforward approach to construct this matrix is to linearize the function $\m h(\cdot)$ around a known operating point, assuming that this yields a detectable pair of $(\mA,\mC)$. 
	This technique allows \eqref{eq:state_space_gen3_lip} to be modified into
	\begin{subnumcases} {\label{eq:state_space_gen2_lip}}
	\dot{\m{x}}= \mA \m x + \m{f}(\m{x},\m{u}) +\m{B_w} \m{q_w} \label{eq:state_space_gen2_lip_a}\\ 
	\m{y}=\mC \m x + \m{h_\mathrm{l}}(\m{x},\m{u}) + \m{D_u} \m u + \m{D_{w}} \m{q_w}. \label{eq:state_space_gen2_lip_b}
	\end{subnumcases}
	Realize that this modification does not change the dynamics of the generator, since \eqref{eq:state_space_gen3_lip} and \eqref{eq:state_space_gen2_lip} are equivalent.
	{\color{black}By expressing \eqref{eq:state_space_gen3_lip_b} as \eqref{eq:state_space_gen2_lip_b}, the generator's output equation can be expressed into linear and nonlinear parts, which allows us to design observer for such nonlinear systems in a linear fashion.} 
	To begin with the observer design, let $\hat{\m x}(t):=\hat{\m x}$ be the estimated state vector and $\hat{\m y}(t):=\hat{\m y}$ be the estimated PMU measurements. 
The proposed observer dynamics are given as
	\begin{subnumcases} {\label{eq:nonlinear_observer_dynamics}}
	\dot{\hat{\m{x}}}= \mA \hat{\m x} + \m{f}(\hat{\m{x}},\m{u}) + \m{B_w} \m r +\mL(\m y-\hat{\m y})\\ 
	\hat{\m{y}} = \mC \hat{\m x} + \m{h_\mathrm{l}}(\hat{\m{x}},\m{u}) + \m{D_u} \m u  + \m{D_w} \m r, 
	\end{subnumcases}
	where $\mL(\m y-\hat{\m y})$ is the Luenberger-type correction term with $\mL\in\mathbb{R}^{4\times 2}$ as a matrix variable; $\m r = \bmat{r_1 &r_2}^{\top}$ where $r_1$ and $r_2$ are any scalars that reflect reasonable ranges of $T_{\mathrm{m}}$ and $E_{\mathrm{fd}}$. 
Defining the estimation error as $\m e(t)  := \m x (t)- \hat{\m x}(t)$, the error dynamics can be computed as 
	\begin{align}
	\dot{\m e}  &= \left(\mA-\mL\mC\right)\m e +\Delta \m f  -\mL\Delta \m{h_\mathrm{l}} + \left(\m {B_{{w}}}-\mL\m {D_{{w}}}\right)\m w, \label{eq:est_error_dynamics}
	\end{align}
	where $\Delta \m f:= \m f(\m x, \m u)-\m f(\hat{\m x}, \m u)$, $\Delta \m{h_\mathrm{l}}:= \m{h_\mathrm{l}}(\m x, \m u)-\m{h_\mathrm{l}}(\hat{\m x}, \m u)$, and $\m w(t) = \m{q_w}(t)-\m r$. Since $\m{q_w}(t)$ is a time-varying signal with unknown values and $\m r$ is fixed, $\m w(t)$ can be perceived as the disturbance acting on  estimation error dynamics \eqref{eq:est_error_dynamics}.

\vspace*{-0.3cm}
\subsection{$\mathcal{L}_{\infty}$ Stability and Observer Synthesis}
	
This section presents the preliminary background and the main contribution of the paper---the robust $\mathcal{L}_{\infty}$ observer design.	The notion of $\mathcal{L}_{\infty}$ stability was first introduced in \cite{pancake2002analysis} for state-feedback control of polytopic systems and then expanded for feedback control of linearized power network models in our recent work~\cite{taha2018robust}. To proceed, we define the $\mathcal{L}_{\infty}$ space as  $\mathcal{L}_{\infty} := \{\m v:[0,\infty)\rightarrow \mathbb{R}^n\,|\,\norm{\m v(t)}_{\mathcal{L}_{\infty}}< \infty\}$ where the $\mathcal{L}_{\infty}$ norm of a signal $\m v(t)$ is defined as $\norm{\m v(t)}_{\mathcal{L}_{\infty}} := \sup_{t \geq 0} \,\norm{\m v(t)}_2$. In other words, $\mathcal{L}_{\infty}$ space defines the space of all bounded functions. The $\mathcal{L}_{\infty}$ observer assumes that $\m w (t)\in \mathcal{L}_{\infty}$. The definition of $\mathcal{L}_{\infty}$ stability with performance level $\mu$ for the estimation error dynamics expressed in the form of \eqref{eq:est_error_dynamics} is given below. 

	\begin{mydef}\label{L_inf_stability}
		Let $\m e(t)\in\mathbb{R}^4$ be the estimation error and $\m z (t)\in \mathbb{R}^4$ be the error performance output defined as $\m z(t) = \mZ \m e(t)$ for a user-defined performance matrix $\mZ\in\mathbb{R}^{4\times 4}$. The nonlinear dynamics \eqref{eq:est_error_dynamics} is  $\mathcal{L}_{\infty}$ stable with performance level $\mu$ if (a) for any bounded disturbance $\m w \in \mathcal{L}_{\infty}$ and zero initial error $\m e_0 = 0$,  $\norm{\m z}_2 \leq \mu \norm{\m w}_{\mathcal{L}_{\infty}}$, (b) there exists $\beta:\mathbb{R}^4\times \mathbb{R}_+\rightarrow \mathbb{R}_+$ such that for any initial error $\m e_0$ and any bounded disturbance $\m w \in \mathcal{L}_{\infty}$, $\norm{\m z}_2 \leq\beta\left(\m e_0,\norm{\m w}_{\mathcal{L}_{\infty}}\right)$, and (c) for any initial error $\m e_0$ and any bounded disturbance $\m w \in \mathcal{L}_{\infty}$, we have $\lim_{t\to\infty}\sup\,\norm{\m z}_2 \leq\mu\norm{\m w}_{\mathcal{L}_{\infty}}$.
	\end{mydef}
	The notion of $\mathcal{L}_{\infty}$ stability with performance level $\mu$ for error dynamics \eqref{eq:est_error_dynamics} can be interpreted as follows. 
	When the initial error is equal to zero, the norm of performance vector $\m z(t)$ for any $t \geq t_0$ is guaranteed to be no more than a scalar multiple of the worst case disturbance, i.e., $\norm{\m z}_2\leq \mu\norm{\m w}_{\mathcal{L}_{\infty}}$. However, if the initial error is nonzero, the norm of performance vector $\m z(t)$ will evolve such that it will not exceed the value of $\mu\norm{\m w}_{\mathcal{L}_{\infty}}$. We also have the norm of performance vector $\m z(t)$ to be bounded by a function of initial condition $\m e_0$ and worst case disturbance $\norm{\m w}_{\mathcal{L}_{\infty}}$. 
	Notice that when $\mu = 0$, the performance vector $\m z (t)$ is irrelevant to the disturbance $\m w(t)$. In contrast, large $\mu$ implies that $\m z (t)$ could be very affected by $\m w(t)$ in the worst case. Thus, when using $\mathcal{L}_{\infty}$ stability for observer design, we want to have performance index $\mu$ to be as small as possible.
	Having derived the error dynamics, we propose a sufficient condition to synthesize the $\mathcal{L}_{\infty}$ observer for \eqref{eq:state_space_gen}; Appendix~\ref{app:l_inf_theorem} has the proof.
	
	\begin{theorem}\label{l_inf_theorem}
		Consider the nonlinear generator and PMU model \eqref{eq:state_space_gen2_lip} and observer \eqref{eq:nonlinear_observer_dynamics} where $\m x,\hat{\m x}\in\mathbfcal{X}$, $\m u\in\mathbfcal{U}$, and the nonlinear functions $\m f:\mathbb{R}^4\times\mathbb{R}^2\rightarrow \mathbb{R}^4$ and $\m {h_{\mathrm{l}}}:\mathbb{R}^4\times\mathbb{R}^2\rightarrow \mathbb{R}^2$ satisfy~\eqref{eq:lipschitz_constants_all}
		 with Lipschitz constants $\gamma_f$ and $\gamma_l$. If there exist $\mP\in\mathbb{S}^4$, $\mY\in\mathbb{R}^{4\times 2}$, and $\m\nu\in\mathbb{R}^6$ so that the following optimization problem is solved
		\begin{subequations}\label{eq:l_inf_theorem}
			\begin{align}
			&J^* = \minimize_{\m P, \m Y, \m\nu} \quad \nu_1\nu_2 + \nu_3 \label{eq:l_inf_theorem_0}\\
			&\subjectto \nonumber \\
			&\hspace{-0.3cm}\bmat{ \mQ +\nu_5\gamma_f^2\mI+\nu_6\gamma_l^2\mI &*&*&*\\
				\mP & -\nu_5\mI&*&*\\
				-\mY^{\top} &\mO &-\nu_6\mI&*\\
				\m {B_{w}}^{\top}\mP-\m {D_{{w}}}^{\top}\mY^{\top}&\mO&\mO&-\nu_4\nu_1\mI} \preceq 0 \label{eq:l_inf_theorem_1}\\
			&\hspace{-0.3cm}\bmat{-\mP & * & * \\
				\mO & -\nu_3\mI & *\\
				\mZ & \mO & -\nu_2\mI}\preceq 0, \;\mP \succ 0, \;\nu_4 > 0 ,\nu_{1,2,3,5,6} \geq 0,\label{eq:l_inf_theorem_2} 
			\end{align}
		\end{subequations}
where $\m Q = \mA^{\top}\mP + \mP\mA -\mC^{\top}\mY^{\top}-\mY\mC +\nu_4\mP$, then  \textit{(i)} the observer gain is $\mL = \mP^{-1}\mY$ and \textit{(ii)} the error dynamics \eqref{eq:est_error_dynamics} are $\mathcal{L}_{\infty}$ stable with performance level $\mu^* = \sqrt{\nu_1^* \nu_2^* +\nu_3^* }$ where the optimal objective value is $J^* = \mu^{*2}$.
	\end{theorem}

We note the following about  $\mc{L}_{\infty}$ observer design problem~\eqref{eq:l_inf_theorem}. First, Theorem \ref{l_inf_theorem} provides a method to obtain the robust gain $\m L$ for observer dynamics \eqref{eq:nonlinear_observer_dynamics} such that the estimation error dynamics \eqref{eq:est_error_dynamics} is robust against the disturbance $\m w(t)$ in the $\mathcal{L}_{\infty}$ sense according to Definition \ref{L_inf_stability}. This result yields a theoretical performance and convergence guarantees for the state estimation error performance output $\norm {\m z(t)}_2$, or $\norm {\m e(t)}_2$, bounding it by $\mu \norm{\m w}_{\mathcal{L}_{\infty}}$. In short, the estimation performance output is guaranteed to lie in a tube of radius $\mu \norm{\m w(t)}_{\mathcal{L}_{\infty}}$ centered around the origin. The case studies in Section~\ref{sec:simulations} examine the applicability of this theoretical bound under different conditions.
    
Second, the $\mc{L}_{\infty}$ observer design problem~\eqref{eq:l_inf_theorem} relies on the computation of Lipschitz constants $\gamma_{f,h}$. These constants define the operational range of the generator defined through $\mathbfcal{X}$ and $\mathbfcal{U}$. The operator can choose to expand the spaces $\mathbfcal{X}$ and $\mathbfcal{U}$ to account for anomalies, deviations from nominal values, faults, and so on. This will impact the Lipschitz constant computation. However, and as we have observed in~\cite{Nugroho2019}, smaller computational Lipschitz constants can be used without compromising the feasibility of observer design problems. Furthermore, and while analytical computation of the Lipschitz provide loose approximation of the Lipschitz constant compared to computational ones~\cite{Nugroho2019}, numerically computed Lipschitz constants can be used instead of analytical ones.

\section{Bounding the Optimal Solution of \eqref{eq:l_inf_theorem}} \label{sec:bounding_SDP}  

 In this section, we discuss the upper and lower bounds on the optimal solution of~\eqref{eq:l_inf_theorem}, seeing that Problem \eqref{eq:l_inf_theorem} is nonconvex due to the bilinear matrix inequality (BMI) constraints appearing in  \eqref{eq:l_inf_theorem_0} and \eqref{eq:l_inf_theorem_1}. The nonconvexity takes a shape of matrix-scalar variable products. Several approaches can be considered to solve nonconvex problems with BMI constraints.
 
 One simple approach is to convert it into a convex SDP by pre-selecting values for $\nu_4$ and either $\nu_1$ or $\nu_2$. This way, a solution can be obtained to the now-convex problem thereby  providing an \textit{upper bound}, namely $\bar{J}$, on $J^* = \mu^{*2}$ which is the optimal objective function value of the nonconvex problem \eqref{eq:l_inf_theorem}. For consistency, the performance index obtained here is denoted by $\bar{\mu}$. If the resulting objective function value is not small enough (big value of $\mu$ translates to a poor performance index), the objective function can be improved by utilizing \textit{successive convex approximations} (SCA) \cite{dinh2012combining}. The idea of SCA is to solve a series of convex problems starting form a feasible initial point. These convex problems, which consist of LMIs, are obtained by linearizing the BMIs using the first-order Taylor approximation. 
 
 On the other hand, a \textit{lower bound} for the nonconvex problem \eqref{eq:l_inf_theorem} can be obtained by first expressing the BMI constraints as \textit{rank-one} equivalent constraints with some other LMI constraints. Since rank-one constraints are also nonconvex, neglecting it yields a convex SDP which then provides a lower bound on $J^*$. This approach is referred to as \textit{SDP relaxation}; see \cite{Taha2018sensor}. The following theorem delineates this approach.  
 
 \begin{theorem}\label{l_inf_relax}
 An SDP relaxation and a lower bound $\underline{J}$ on $J^*$ of the nonconvex problem \eqref{eq:l_inf_theorem} can be obtained by solving the following \textbf{convex} SDP with additional variables $\m \Xi \in \mathbb{S}^4$, $\m\Psi_{i,j}\in\mathbb{S}^3$ for all $i,j$, $\m \Phi\in\mathbb{S}^3$, $\m \Theta\in\mathbb{S}^3$, $\lambda\in \mathbb{R}$, and $\sigma\in \mathbb{R}$
 		\begin{subequations}\label{eq:l_inf_relax}
 		\begin{align}
 		\barbelow{J} &= \minimize_{\substack{\m P, \m Y, \m\nu, \m \Xi, \m\Psi \\ \m \Phi, \m \Theta, \lambda, \sigma}} \quad \lambda + \nu_3 \label{eq:l_inf_relax_0}\\
 		&\subjectto \;\;
 		E(\m P, \m\nu, \m \Xi, \m\Psi_{i,j},\m \Phi, \m \Theta, \lambda, \sigma) = 0 \label{eq:l_inf_relax_1}\\ 	
 		&\qquad \qquad L(\m P, \m Y, \m\nu, \m \Xi, \m\Psi_{i,j},\m \Phi, \m \Theta, \lambda, \sigma) \preceq 0, \label{eq:l_inf_relax_2}
 		\end{align}
 	\end{subequations}
where \textit{(i)} $E(\cdot)$ and $L(\cdot)$ are convex matrix equality and inequality constraints and \textit{(ii)} it holds that $\barbelow{J} = \lambda + \nu_3 \leq \mu^{*2} = J^*$ in which $\mu^* = \sqrt{\nu_1^*\nu_2^* + \nu_3^*}$ is obtained from solving \eqref{eq:l_inf_theorem}. 
 \end{theorem}  
\noindent The proof is included in Appendix \ref{app:theorem_SDP_relax} which also provides the closed form expressions of \eqref{eq:l_inf_relax_1} and \eqref{eq:l_inf_relax_2}. 

As \eqref{eq:l_inf_relax} provides a lower bound, it is useful for assessing the quality of solution from solving \eqref{eq:l_inf_theorem} either by fixing some constants apriori or using SCA---and hence assessing the quality of the upper bound. 
Furthermore, the SDP relaxation in \eqref{eq:l_inf_relax} can be efficiently solved \textit{optimally} using any convex optimization solver. This is in contrast to the nonconvex problem~\eqref{eq:l_inf_theorem} which is much harder to solve. 
We note that, however, for this lower bound $\underline{J}$ to be useful, its value needs to be not too distant from an upper bound $\bar{J}$, thereby guaranteeing that $ \underline{J} \leq J^* \leq \bar{J} $. The quality of these theoretical bounds are thoroughly investigated in Section \ref{sec:simulations}.

{\color{black}
\section{Case Studies}~\label{sec:simulations}

We test the proposed approach on the 16-machine, 68-bus system extracted from the PST toolbox \cite{chow1992toolbox,PST_manual}. The same test system has been used in \cite{access_18}. 
The input vector $\m{u}(t)$, which consists of $i_{\mathrm{R}}$ and $i_{\mathrm{I}}$, and unknown input $\m {q_m}(t)$, which consists of $T_{\mathrm{m}}$ and $E_{\mathrm{fd}}$, are obtained from simulations of the whole system in which each generator is using a transient model with IEEE Type DC1 excitation system and a simplified turbine-governor system \cite{report1981excitation,access_18}. 
To generate dynamic response, we apply a three-phase fault at Bus 32 of Branch 32-30 and clear the fault at the near and remote ends after 0.05 and 0.1 seconds. All DSE simulations are performed for the pre, during, and post-contingency system within 15 seconds time frame. 

We consider that this generator is connected to a PMU with sampling rate of 60 frames/s. To obtain $\mathbfcal{X}$ and $\mathbfcal{U}$ which contain upper and lower bounds on the state and input (namely $\m x_{\mathrm{min}}$, $\m x_{\mathrm{max}}$, $\m u_{\mathrm{min}}$, and $\m u_{\mathrm{max}}$), their minima and maxima inside this time frame are measured. 
All of the following simulations are performed using MATLAB R2016b running on a 64-bit Windows 10 operating system with 2.5GHz Intel\textsuperscript{R} Core\textsuperscript{TM} i7-6700 CPU and 16 GB of RAM.
We use YALMIP \cite{Lofberg2004} as the interface and MOSEK \cite{andersen2000mosek} solver to solve the SDPs corresponding to the $\mc{L}_{\infty}$ observer. It is worth mentioning that the generator response as well as the estimation via observer are all simulated using MATLAB's \texttt{ode45} with its default settings.

\subsection{Single Generator DSE with $\mc{L}_{\infty}$ Observer}\label{ssec:test1}
 
For the \nth{4}-order model from Section~\ref{sec:model}, in this numerical test we consider Generator 4 with initial condition  $\m x_{0} = [1.1548\quad \omega_0 \quad 0.8632 \quad 0.6222]$, which is obtained from PST. Realize that the \nth{4}-order model does not include any differential equations that represent generator's controller. Thus, in the following simulations we use unstable generator trajectories to evaluate the performance of the observer on tracking such response (a higher order generator model is also tested in subsequent sections).
Four different cases based on the \nth{4}-order model are investigated. Table \ref{tab:comp} lists these cases.

\subsubsection{Case 1}	

Here the generator is only subject to unknown inputs $\m {q_m}(t)$. To obtain the observer gain matrix \eqref{eq:l_inf_theorem} is solved as LMIs by simply fixing $\nu_4 = 1$ and $\nu_2 = 50$, as discussed in Section \ref{sec:bounding_SDP}. \textcolor{black}{The performance matrix is chosen to be $\mZ = 2\times 10^{-4}\mI$, whereas the matrix $\mC$ is set to be equal to $10D_{\m x}\m h$, where $D_{\m x} \m h$ is the Jacobian matrix of $\m h(\cdot)$, linearized around $\check{\m x}$ and $\check{\m u}$ where $\check{\m x} = (\m x_{\mathrm{min}} + \m x_{\mathrm{max}})/2$ and $\check{\m u}=(\m u_{\mathrm{min}} + \m u_{\mathrm{max}})/2$. This setup gives an observable pair $(\mA,\mC)$.  Changing this operating point does \textit{not} adversely impact the results.} The Lipschitz constants $\gamma_f$ and $\gamma_h$ are obtained from using \cite[Theorem 1]{Nugroho2019}, where the Lipschitz constant $\gamma_l$ can be computed as $\gamma_l = \gamma_h + \norm{\mC}_2$. This returns $\gamma_f = 379.1$ and $\gamma_l = 30.1$. We set $r_1 = 0.9$ and $r_2 = 2.1$ which are both different from the steady-state value of $T_{\mathrm{m}}$ and $E_{\mathrm{fd}}$ 
that are approximately $0.79$ and $1.88$ as shown in Fig. \ref{fig:UI_ori}. 
The observer's initial condition is randomly chosen as $\hat{\m x}_{0} = [0.7548\quad  \omega_0 \quad 1.3632 \quad 0.8222]$, which is significantly different from that of the generator $\m x_0$. The observer is \textit{not} sensitive to initial conditions.

After solving Problem \eqref{eq:l_inf_theorem}, we obtain a performance index $\bar{\mu} = 4.1904\times 10^{-4}$. 
Fig.~\ref{fig:new_scen_z_all} shows that the estimation error norm $\norm {\m e(t)}_2$ generally converges close to zero---the fluctuations are caused by uncertainty due to unknown inputs.
In addition, we also  observe that the definition of $\mathcal{L}_{\infty}$ stability is indeed satisfied, since the norm of performance output $\norm {\m z(t)}_2$ converges to a value which is less than $\bar{\mu} \norm { \m w(t)}_{\mathcal{L}_{\infty}} = 1.0553\times 10^{-4}$. 
This finding is summarized in Table \ref{tab:comp} and corroborates the performance guarantees from Theorem~\ref{l_inf_theorem}. Note that in Fig.~\ref{fig:new_scen_z_all} the above value is scaled with a factor $k = 5\times 10^{3}$ so that it is proportional with the estimation error norm.
	
\begin{figure}[t]
	\hspace{0.0cm}\centerline{\includegraphics[keepaspectratio=true,scale=0.48]{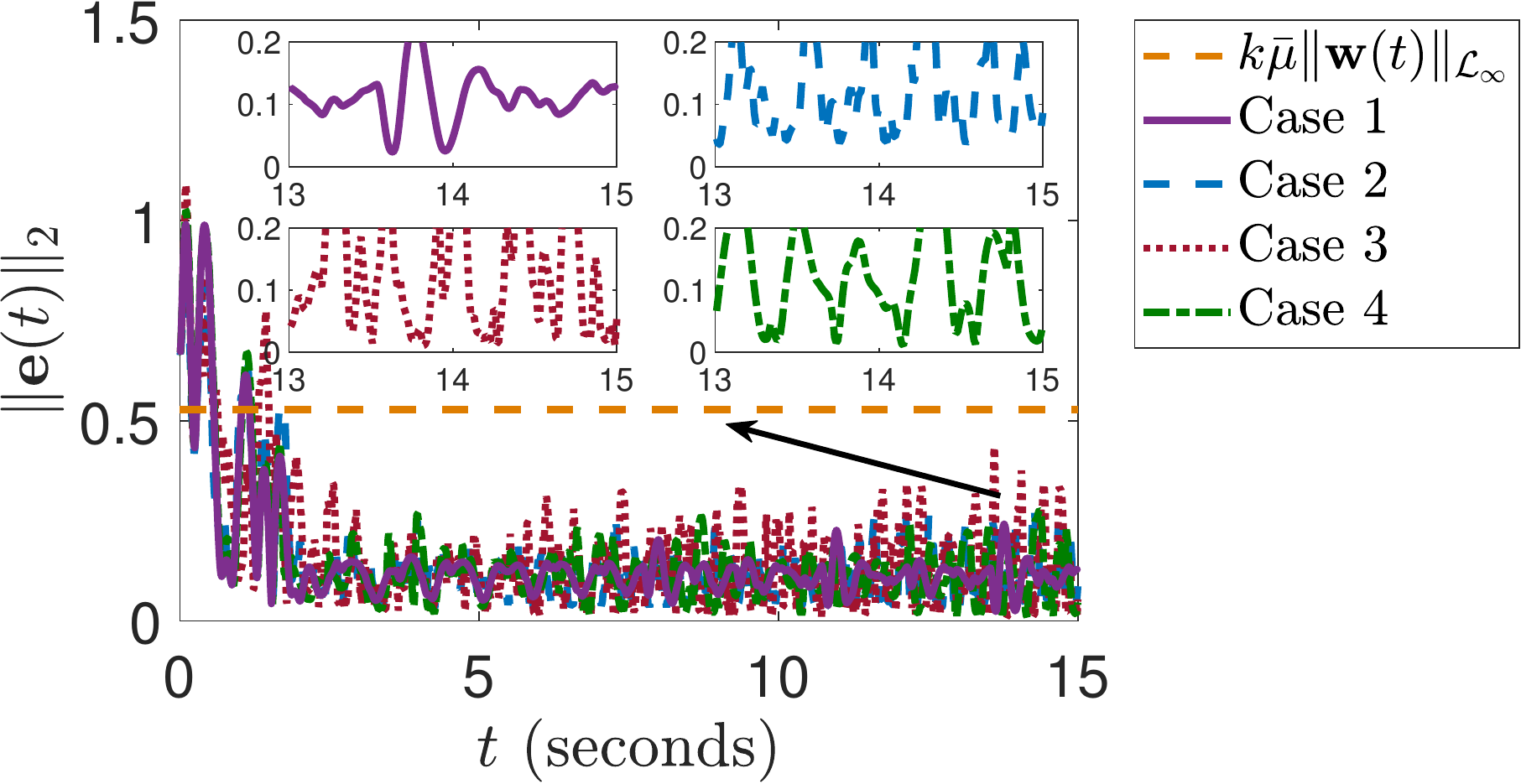}}
	\caption{The estimation error $\Vert\m e(t)\Vert_2$ and scaled worst-case disturbance $k\bar{\mu}\Vert\m w(t)\Vert_{\mathcal{L}_{\infty}} = 0.5276$ for Cases 1--4.} 
	\label{fig:new_scen_z_all}
\vspace{-0.2cm}
\end{figure}

	\begin{figure}[t]	\vspace{-0.0cm}\hspace{0.0cm}\centerline{\includegraphics[keepaspectratio=true,scale=0.4]{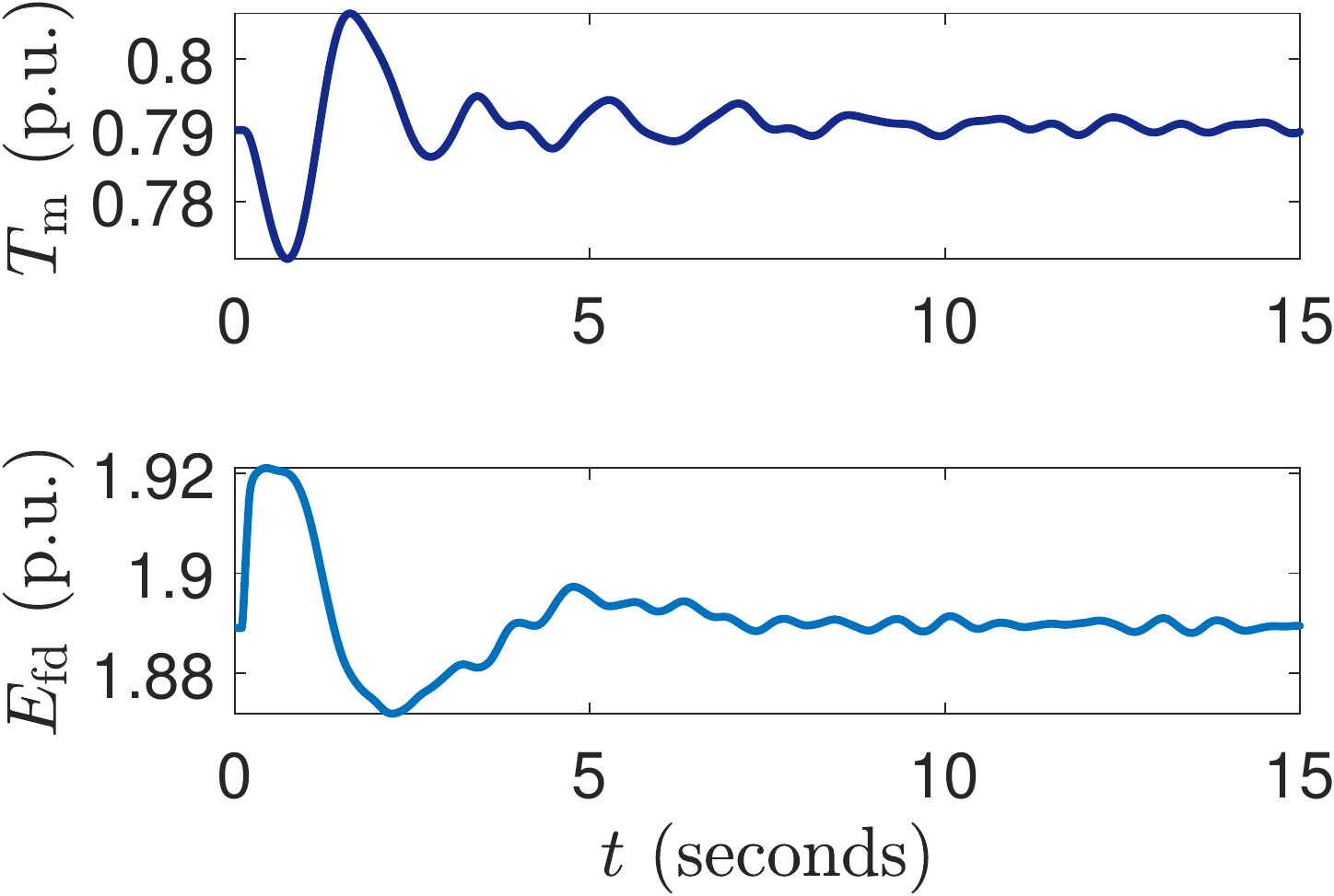}}
	\caption{Trajectories of unknown inputs $T_{\mathrm{m}}$ and $E_{\mathrm{fd}}$ in Cases 1--4. } 
	\label{fig:UI_ori}
	\vspace{-0.2cm}
\end{figure}

	\subsubsection{Cases 2--4}
	Cases 2--4 consider process, input, and measurement noises such that the generator's dynamic model \eqref{eq:state_space_gen3_lip} is expressed as
	\begin{subnumcases} {\label{eq:state_space_gen3}}
	\dot{\m{x}}= \mA \m x + \m{f}(\m{x},\m{u}) +\m{{B}_w} \m{q_w} + \m{v_{p}}\\
	\m{y}=\mC \m x + \m{h_\mathrm{l}}(\m{x},\m{u}) + \m{D_u} \m u  + \m{{D}_w} \m{q_w} + \m{v_{m}}.
	\end{subnumcases}
	In \eqref{eq:state_space_gen3}, the vector $\m{v_{p}}$ represents process noise while $\m{v_{m}}$ represents measurement noise. The noise is clearly not known to the observer hence keeping $\m B_w$ and $\m D_w$ the same as in Case 1.
	For Cases 2--4, the Gaussian process noise is generated assuming diagonal covariance matrix which entries are the square of {\color{black}$5$\%} of the largest state changes. For Case 2, the Gaussian measurement noise covariance matrix is also diagonal with variance $0.05^2$.
	The Laplace noise on the PMU measurement $i$ for Case 3 is characterized by the signal $\m{v}_{mi} = m - s \cdot \mathrm{sgn}(R_1)\cdot\mathrm{ln}(1- 2\vert R_1\vert)$,
	where $m = 0$, $s = 0.02$, and $R_1$ is a number randomly chosen inside $(-0.5,0.5]$. \textcolor{black}{For Case 4, Cauchy noise on each entry $i$ is generated as $\m{v}_{mi} = a+ b \cdot \mathrm{tan}\big(\pi(R_2-0.5)\big)$,
	where $a = 0$,  {\color{black}$b = 1\times 10^{-3}$}, and $R_2$ is a random number inside $(0,1)$. }
	
The motivation for testing non-Gaussian noise is as follows. In \cite{error}, extensive results using field PMU data from
WECC system reveal that the Gaussian assumption for measurement noise is questionable. Therefore, it is necessary to test non-Gaussian noises. 
Here we test Gaussian noise, Cauchy noise, and Laplace noise for the PMU measurements.  
The latter two non-Gaussian noises are suitable for being used to test a robust dynamic state estimator due to the fact that typical noises with heavy-tail distributions are more challenging to deal with than Gaussian noise.

The estimation results are depicted in Fig. \ref{fig:scen2} for Case 3 and Case 4. We observe that the estimated states and outputs still follow the actual ones in spite of the presence of process and measurement noise---the estimation for Case 2 yields similar results. The estimation error norm for these cases are shown in Fig. \ref{fig:new_scen_z_all}. It is worth noting that the definition of $\mathcal{L}_{\infty}$ stability is still satisfied for the three cases. Table \ref{tab:comp} summarizes the comparison of the norm of performance matrix $\norm {\m z(t)}_2$ with $\bar{\mu} \norm { \m w(t)}_{\mathcal{L}_{\infty}}$,  	{\color{black}where we also compare the \textit{Root Mean Square Error} (RMSE) for all cases computed using the following formulation
		\begin{align*}
		\footnotesize
		\mathrm{RMSE} &= \sum_{i=1}^{n} \sqrt{\frac{1}{t_f}\sum_{t=1}^{t_f}\big(x_i(t)-\hat{x}_i(t)\big)^2}.
		\end{align*}
It is later found that the choice of $\m r$  plays a role in error quality, although not to a great extent.} Any $\m r$ that is sufficiently close to the steady state value of unknown inputs will return better estimation results. The next section showcases the performance of the observer under ramp changes in the unknown inputs.
	
\begin{figure}[t]
	\vspace{0.0cm}
	\centering
	\subfloat[]{\includegraphics[keepaspectratio=true,scale=0.38]{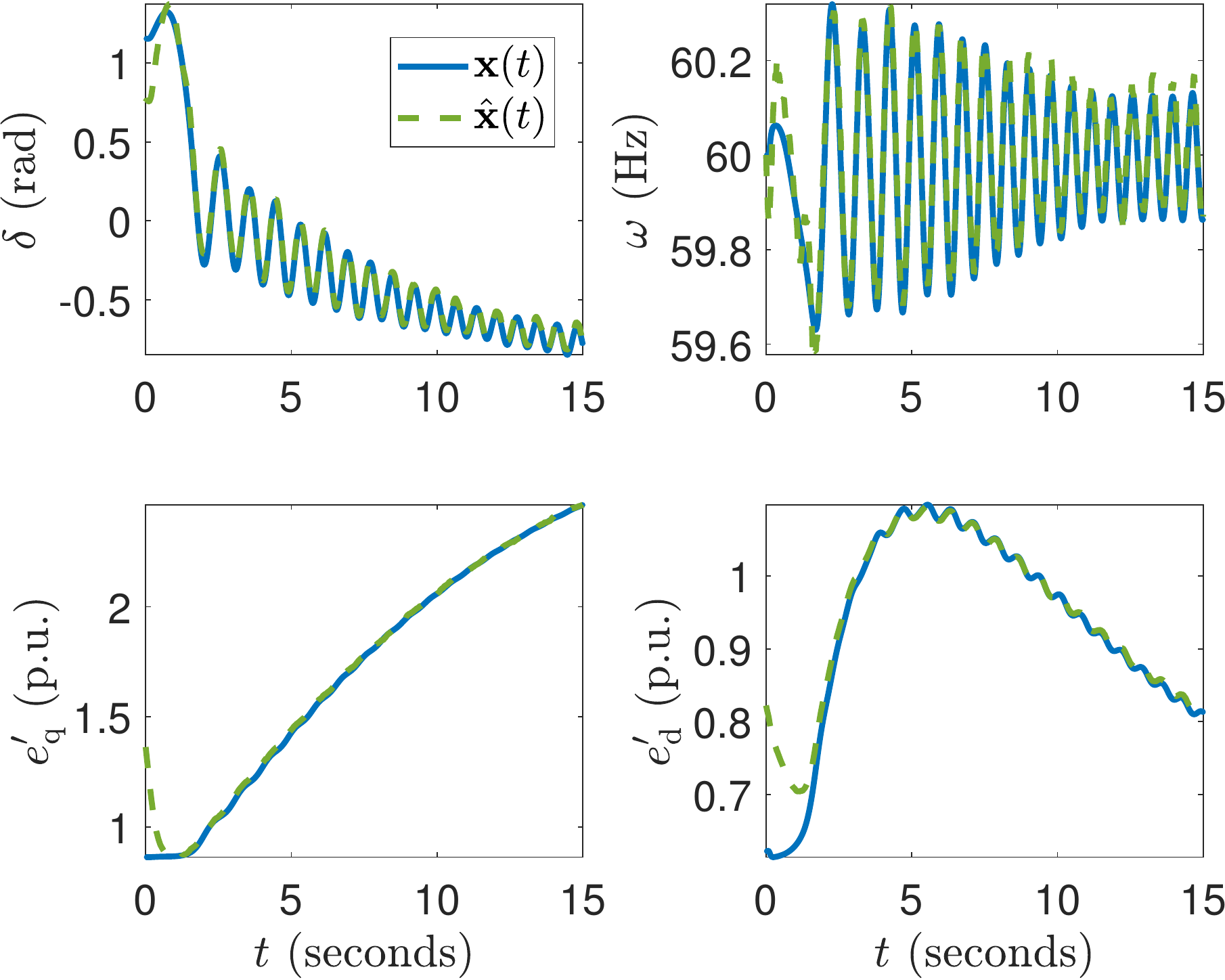}\label{fig:scen2-1}}{\vspace*{-0.15cm}}
	\subfloat[]{\includegraphics[keepaspectratio=true,scale=0.37]{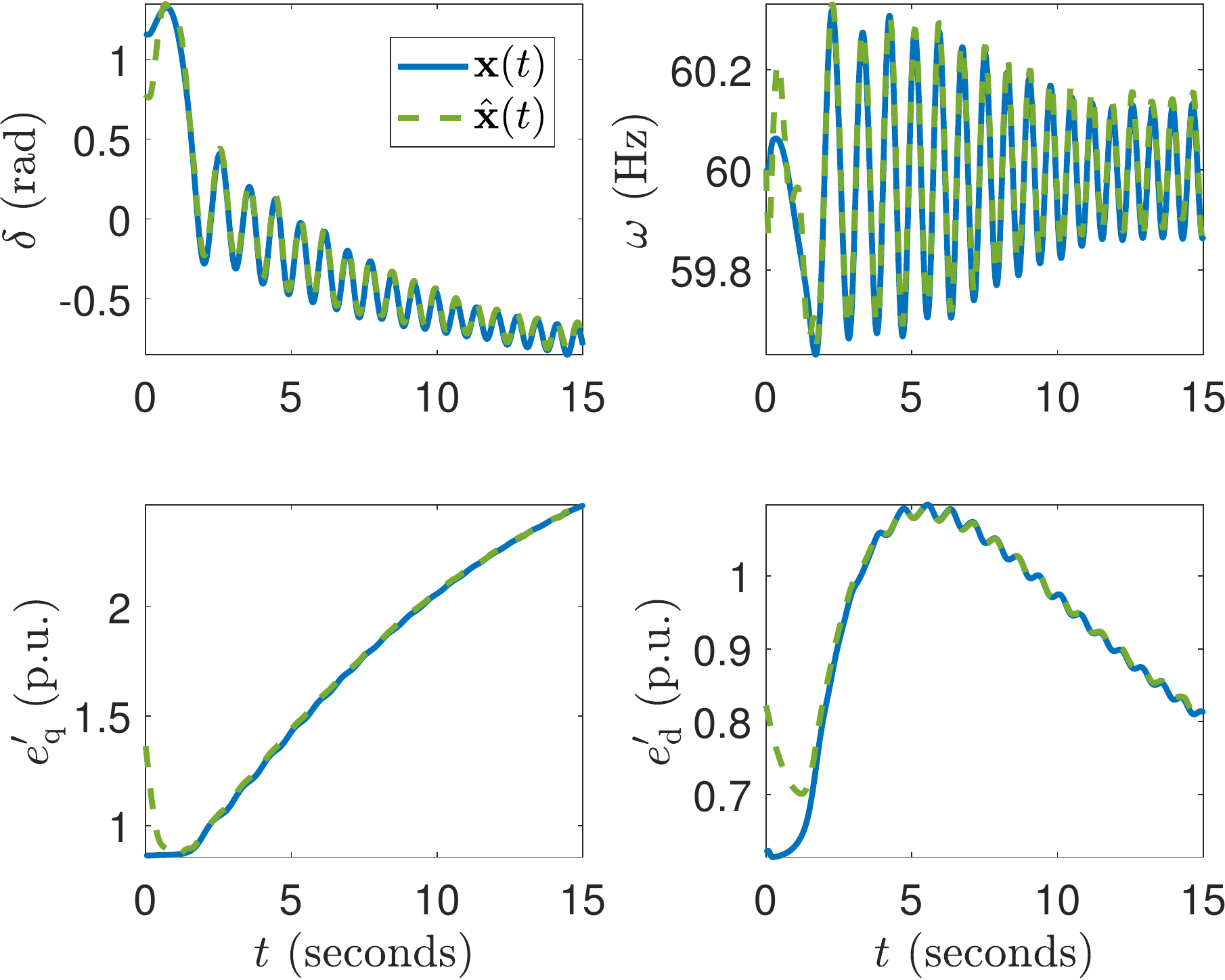}\label{fig:scen2-2}}{	\vspace*{-0.1cm}}
	\noindent \caption{\color{black} Actual and estimated states for (a) Case 3 and (b) for Case 4. }\label{fig:scen2}
\end{figure}

\vspace*{-0.2cm}
\subsection{DSE Under Ramp and Step Changes on $T_{\mathrm{m}}$ and $E_{\mathrm{fd}}$}\label{ssec:test2}

In this numerical experiment, we modify the unknown inputs $T_{\mathrm{m}}$ and $E_{\mathrm{fd}}$, represented by vector $\m {q_w}(t)$, by artificially adding ramp and step changes. Although this does not represent realistic behavior per se, this kind of change can showcase the performance of $\mathcal{L}_{\infty}$ observer amidst sudden changes in the unknown inputs. The resulting unknown input signals are depicted in Fig. \ref{fig:UI_ramp_step}. 
With the modified unknown inputs, we simulate the $\mc{L}_{\infty}$ observer for Case 2 described in Section \ref{ssec:test1}. The results of this experiment are given in Fig. \ref{fig:scen2_ramp_step}. We can see that, albeit the unknown inputs have been modified and while the value for $\m r$ provided to the observer did not change from Case 2, the estimated states and outputs are still on track with generator's actual states and outputs. {\color{black} Notice the large errors at some instances due to these sudden changes and the converging estimated states to the real ones at the other remaining instances.}

	\begin{figure}[t]	\vspace{-0cm}\hspace{0.0cm}\centerline{\includegraphics[keepaspectratio=true,scale=0.35]{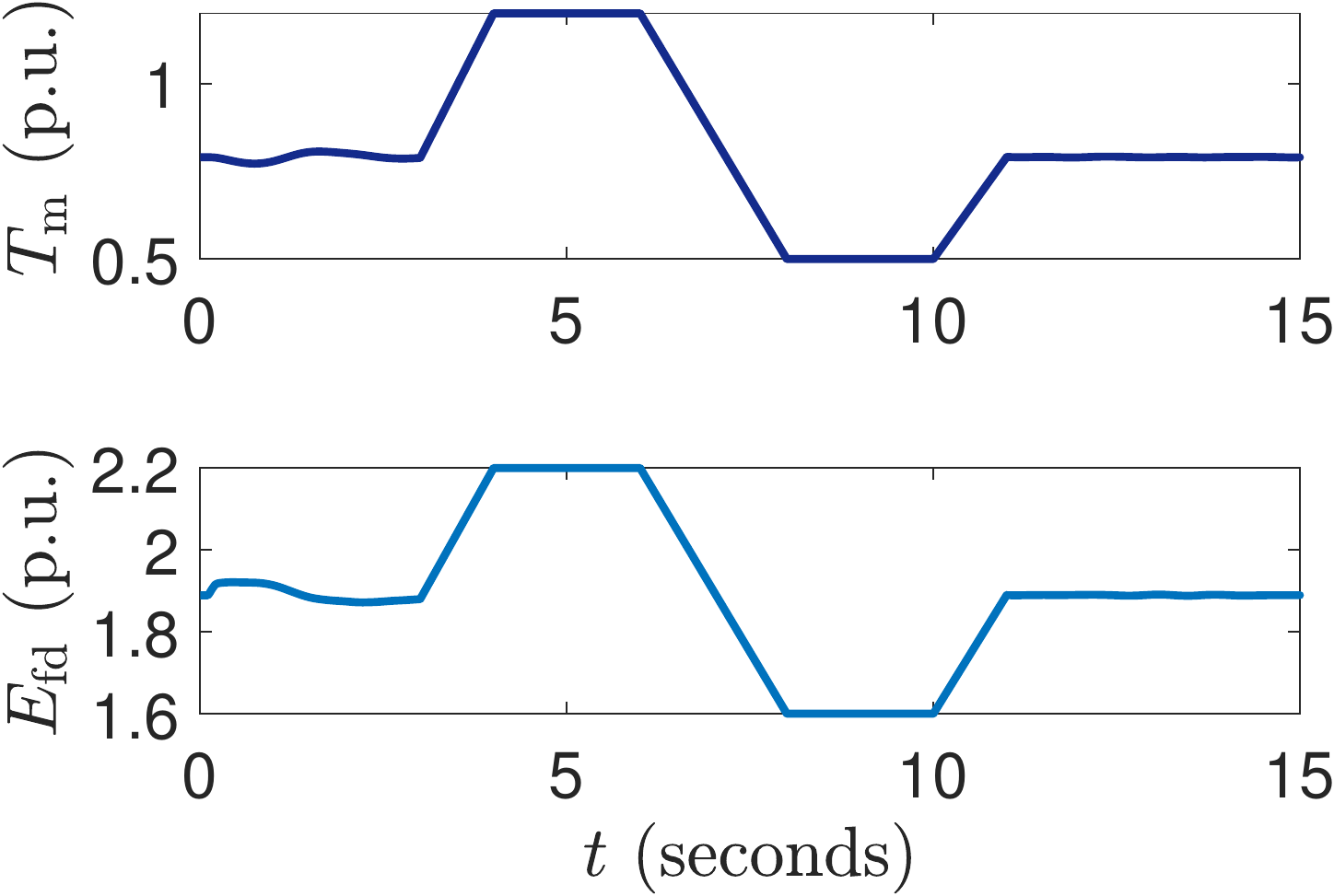}}
	\vspace{-0.5cm} 
	\caption{Trajectories of unknown inputs with ramp and step changes.} 
	\label{fig:UI_ramp_step}
\end{figure}

\begin{figure}[t]
	\vspace{-0.0cm}
	\centering
	{\includegraphics[keepaspectratio=true,scale=0.35]{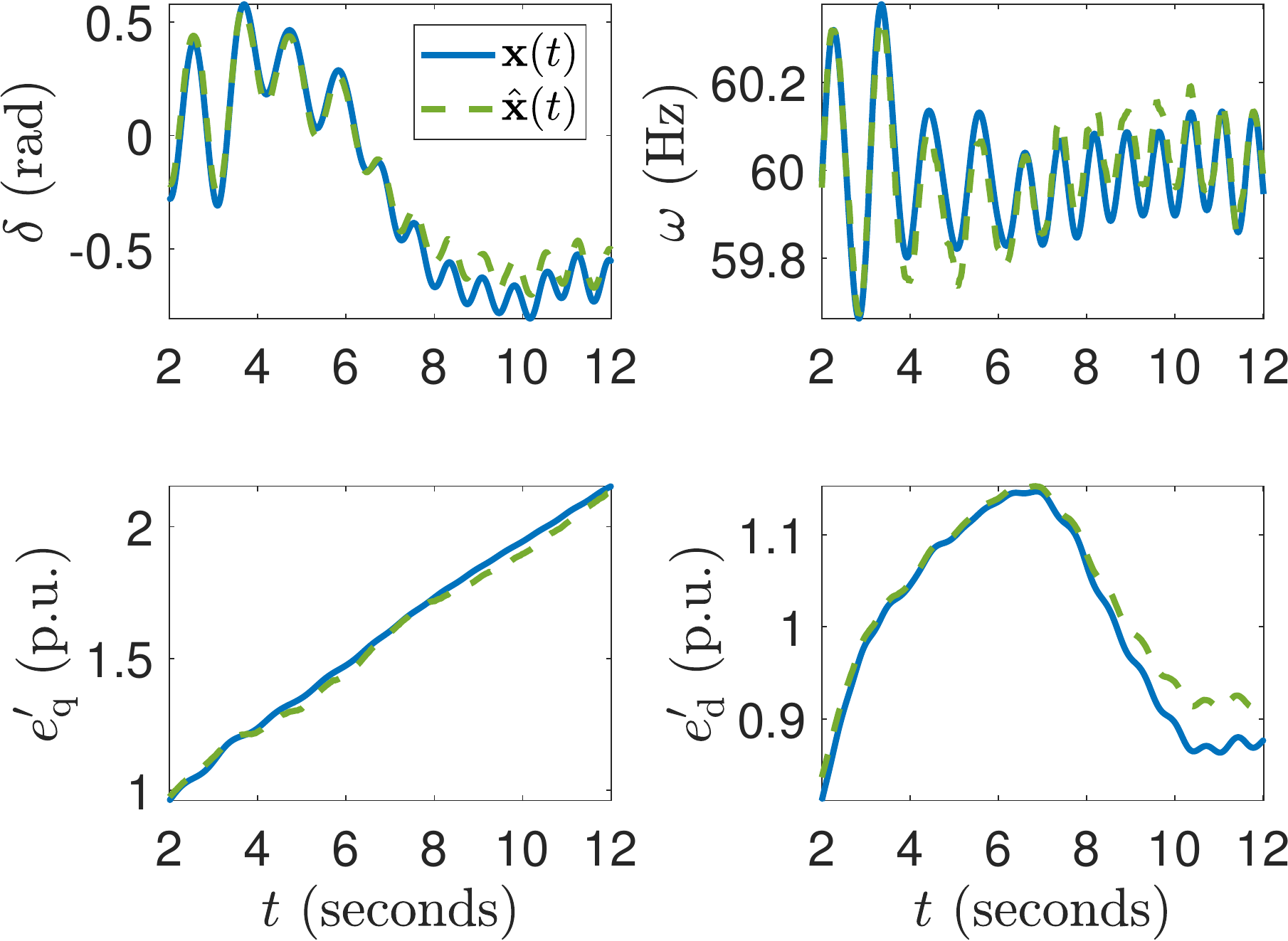}\label{fig:scen2_ramp_step-1}}{\vspace*{-0.1cm}}
	\noindent \caption{\color{black} Actual and estimated states  for Case 2 with ramp and step changes on unknown inputs $\m {q_m}(t)$.}\label{fig:scen2_ramp_step}
\end{figure}

\begin{table*}[t]
	\vspace{-0.3cm}
	\footnotesize
	\centering\color{black}
	\begin{threeparttable}[b]
		\caption{  Numerical Tests Summary for Cases 1--6 
		} 
		\renewcommand{\arraystretch}{1.3}
	\begin{tabular}{c| c | l | l | l |c|l}
	\hline
	Model & Scenario & Description & \multicolumn{1}{l|}{$\mathrm{max}(\Vert\m z(t)\Vert_2)$\tnote{$\dagger$}} & \multicolumn{1}{l|}{$\bar{\mu} \Vert\m w(t)\Vert_{\mathcal{L}_{\infty}}$} & \multicolumn{1}{l|}{STG?\tnote{$\star$} } & \multicolumn{1}{l}{RMSE} \\
	\hline \hline
	\multicolumn{1}{c|}{\multirow{4}[2]{*}{\nth{4}-order}} & \textit{Case 1} & Estimation under unknown inputs $T_{\mathrm{m}}(t)$ and $E_{\mathrm{fd}}(t)$ &   $4.8362\times 10^{-5}$    & $1.0553\times 10^{-4}$      & \checkmark & $0.34208$\\
	& \textit{Case 2} & Case 1 with Gaussian process and measurement noise &  $5.4028\times 10^{-5}$     & $1.0553\times 10^{-4}$      & \checkmark & $0.34211$ \\
	& \textit{Case 3} & Case 1 with Gaussian process noise and Laplace measurement noise &  $8.7081\times 10^{-5}$     &  $1.0553\times 10^{-4}$     & \checkmark & $0.36605$\\
	& \textit{Case 4} & Case 1 with Gaussian process noise and Cauchy measurement noise &  $5.5270\times 10^{-5}$     &  $1.0553\times 10^{-4}$     & \checkmark & $0.35160$\\
	\hline
	\multirow{2}[2]{*}{\nth{10}-order} & \textit{Case 5} &  Case 2 with initial condition $ \hat{\m x}_{0}^{(1)}$   &   $1.1021\times 10^{-5}$    & $1.1295\times 10^{-5}$      & \checkmark & $0.27106$\\
	& \textit{Case 6} & Case 3 with initial condition $\hat{\m x}_{0}^{(2)}$ &  $2.3373\times 10^{-5}$     &  $2.9292\times 10^{-5}$     & \checkmark & $0.28574$\\
	\hline 
\end{tabular}
		\label{tab:comp}%
		\begin{tablenotes}
			\item[$\dagger$] $\mathrm{max}(\Vert\m z(t)\Vert_2)$ denotes the maximum norm of $\m z(t)$ for $t \in [10,15]$, thereby assessing the asymptotic convergence guarantees. 
			\item[$\star$] {Satisfied Theoretical Guarantees} (STG) is checked when $\mathrm{max}(\Vert\m z(t)\Vert_2)\leq \bar{\mu} \Vert\m w(t)\Vert_{\mathcal{L}_{\infty}}$.
		\end{tablenotes} 
	\end{threeparttable}
	\vspace{-0.5cm}
\end{table*}%
\normalcolor

\vspace*{-0.2cm}
\subsection{Robust Observer Performance Under Model Uncertainty}\label{ssec:test3}

This section is dedicated to show the performance of the proposed DSE method under model uncertainty. Here we use Case 2  introduced in Section \ref{ssec:test1} as a benchmark but other cases have shown similar performance. To simulate model uncertainty, we modify the generator dynamics as follows
\begin{subnumcases} {\label{eq:state_space_gen4}}
\dot{\m{x}}= (\mA + \delta\mA)\m x + \m{f}(\m{x},\m{u}) + \delta\m{f}(\m{x},\m{u}) +\m{{B}_w} \m{q_w} + \m{v_{p}}\nonumber\\
\m{y}=\mC \m x + \delta\m{h_\mathrm{l}}(\m{x},\m{u}) + \m{D_u} \m u  + \m{{D}_w} \m{q_w} + \m{v_{m}},\qquad \nonumber 
\end{subnumcases}
where $\delta\m{h_\mathrm{l}}(\m{x},\m{u}) := -\mC \m x + \delta\m{h}(\m{x},\m{u})$ and $\delta\in [0,1]$ is a constant used to determine the magnitude of model uncertainty. With this model, the observer dynamics are left intact, as in the change in the parametric uncertainty is not provided to the estimator dynamics. Three different values of $\delta$, which are $0.02$, $0.05$, and $0.1$, are used to simulate $2$\%, $5$\%, and $10$\% model uncertainty in the generator parameters. Fig. \ref{fig:error_norm_mod_unc} shows the estimation error norm for each percentage of model uncertainty, from which we can see that the steady state error norm is getting bigger as the percentage increases. These results are indeed expected as the discrepancy between the assumed and actual model is nothing but disturbance. We observe the $\mc{L}_{\infty}$ observer is still somewhat tolerating parametric uncertainty as well as other sources of unknown inputs from Case 2. This showcases that the proposed estimator is robust to \textit{minor} parametric uncertainty, albeit it theoretically does not account for it. Further theoretical developments can be pursued to account for major parametric uncertainty, but this is outside the scope of this work.

\vspace*{-0.2cm}
\subsection{Comparing $\mc{L}_{\infty}$ Observer With Other Estimation Methods}
As mentioned earlier, various methods have been developed for DSE in power systems such as EKF, UKF, and Square Root UKF (SR-UKF) 
\cite{ekf1,ghahremani2011dynamic,valverde2011unscented,pwukf1,pwukf2,qi,qi2015dynamic}. Here, we do a comparative study to measure the performance of the $\mc{L}_{\infty}$ observer in comparison with EKF, UKF, and SR-UKF. A brief theoretical background about these KFs along with their set up for this comparison are given as follows.

\noindent	$\bullet$  EKF may be regarded as the simplest KF-based estimator for nonlinear systems. It basically has the same principle as the original KF except that it uses the Jacobian matrix of the nonlinear transformation in KF algorithm. The initial state error covariance matrix for EKF is set to be 
\begin{align}
\m P_0 = \mathrm{diag}\left({\bmat{ \tfrac{\pi}{90}& 2\times 10^{-3} \times 60 \pi &10^{-3}&10^{-3}}}\right). \label{eq:init_err_cov}
\end{align}

\noindent	$\bullet$  In contrast to EKF, UKF does not require linearization of the dynamics. Based only on the nonlinear process/measurement models, it uses an \textit{unscented transformation} to extract and  estimate the mean and covariance data that have gone through nonlinear transformations. In UKF, sigma points are generated to represent Gaussian distribution. The constants to generate sigma points are set to be $\alpha = 0.1$, $\beta = 2$, and $\kappa = -1$. The initial state error covariance matrix is the same as in \eqref{eq:init_err_cov}. 

\noindent	$\bullet$  SR-UKF is an improved variant and more numerically stable than UKF \cite{qi2015dynamic}. The SR-UKF is set to use the same constants as those on UKF to generate sigma points. The square root of the initial state error covariance matrix is $\m S_0 = \sqrt{\m P_0}$.

As EKF, UKF, and SR-UKF estimate in discrete-time, the discrete-time model of the generator is obtained based on the \nth{2}-order Taylor approximation of generator dynamics \eqref{eq:state_space_gen}. Furthermore, the term  $\m{B_w} \m r $ is appended to all of these stochastic estimator process dynamics, thereby ensuring fairness when comparing them with the robust observer. 

In this comparison two scenarios are considered. The first scenario is based on Case 2 (see Section \ref{ssec:test1}), in which the generator is subject to Gaussian process and measurement noise and generator unknown inputs. In the second scenario, we use Case 3 as discussed in Section \ref{ssec:test1} but with the addition of $2\%$ model uncertainty. The process and measurement noise covariance matrices for both scenarios are the same as the ones in Case 2. The numerical test results for the first scenario are illustrated in Figs. \ref{fig:new_scen2_KF} and  \ref{fig:scen2_KF}. 
We particularly observe from these figures that the observer is able to estimate generator's state with relatively very small errors. Albeit SR-UKF can perform an adequate estimation, the observer's performance is better. On the other hand, EKF and UKF give relatively worse estimation than SR-UKF and the observer. This observation is also supported from the plots of estimation error norm given in Fig. \ref{fig:new_scen2_KF}. From this figure we see that the $\mc{L}_{\infty}$ observer can perform DSE with the smallest steady state error norm compared to EKF, UKF, and SR-UKF. Other test cases show a similar trend. 

Interestingly, the $\mc{L}_{\infty}$ observer is also performing better than the other methods in the second scenario (discussed above), as shown in Fig. \ref{fig:new_scen2_KF_laplace}. 
In order to obtain a clear quantitative comparison, we compute their respective RMSE---see Table \ref{tab:computational_time_KF} for the corresponding results.  
We observer that the observer returns the smallest RMSE, in comparison with the other KF-based methods.

In addition to RMSE comparison, we also find that the observer is actually more computationally efficient in performing DSE than EKF, UKF, or SR-UKF. This is corroborated from numerical data shown in Table \ref{tab:computational_time_KF}, in which the $\mc{L}_{\infty}$ observer has the least simulation running time, followed by UKF and SR-UKF, while EKF being the least efficient. 
The reason why $\mc{L}_{\infty}$ observer has the least running time is due the fact that it only utilizes a simple one-step predictor with fixed gain matrix, as opposed to KF-based estimators; see the observer dynamics~\eqref{eq:nonlinear_observer_dynamics}. From Table \ref{tab:computational_time_KF}, one can also note that  EKF is less efficient than UKF and SR-UKF, which contradicts the knowledge that EKF has smaller computational complexity than UKF and SR-UKF. We presume that this is caused by implementation issues on MATLAB, and is not representative of what happens at scale when a multi-machine network model is considered with hundreds of generators. 
\begin{table}[t]
	\vspace{-0.1cm}
\color{black}	\footnotesize	\centering \renewcommand{\arraystretch}{1.3}
	\begin{threeparttable}[b]
\caption{Comparison of Simulation Running Time $\Delta t$ (s)  and RMSE for Observer, EKF, UKF, and SR-UKF}
\begin{tabular}{l|l|l|l|l}
	\hline
	\multirow{2}{*}{DSE Method} & \multicolumn{2}{c|}{Case 2} & \multicolumn{2}{c}{Case 3 + $2\%$ uncertainty} \\ \cline{2-5} 
	&   $\Delta t$ (s)        &   RMSE        &     $\Delta t$ (s)        &     RMSE      \\ \hline\hline
	$\mc{L}_{\infty}$ \textit{Observer}\tnote{$\dagger$}	&    $20.91 $      &   $ 0.3489 $      &   $25.74$        &      $0.3742$     \\ \hline
	\textit{EKF}	&  $ 74.01 $       &    $2.7196$       &   $73.75$        &     $2.6096$      \\ \hline
	\textit{UKF}	&   $ 40.33 $      &    $2.7207$       &   $40.15$        &     $2.6111$      \\ \hline
	\textit{SR-UKF}	&    $ 40.33$     &    $0.6092$       &   $40.12$        &    $0.6256$       \\ \hline
\end{tabular}\label{tab:computational_time_KF}
	\begin{tablenotes}
	\item[$\dagger$] It took $1.2641\sec$ to obtain the observer gain matrix. This is not included in the $20.91\sec$ figure shown in the table, seeing that this computation is offline. However, this does not change the main takeaway. 
\end{tablenotes} 
\end{threeparttable}
\end{table}
\vspace*{-0.4cm}

\vspace*{-0.0cm}
\subsection{Extension to \nth{10}-Order Generator Model}\label{ssec:test4}

In the previous sections, we use a \nth{4}-order transient model to perform DSE whereby the generator's controllers are not modeled. This allowed us to test the estimator's performance during unstable  transients. In order to further test the proposed $\mathcal{L}_{\infty}$ robust observer on higher-order generator models under stable conditions, in this section we use a \nth{10}-order generator model for DSE with a transient generator model, IEEE Type DC1 excitation system, a simplified steam turbine governor system \cite{report1981excitation,sauer1998power}. We reformulate the \nth{10}-order model taken from \cite{sauer1998power} such that with process and measurement noise, it can be expressed as
 \begin{subnumcases} {\label{eq:state_space_gen_10_ord}}
\dot{\m{x}}= \mA \m x + \m{f}(\m{x},\m{u}) + \m{{B}_w} \m{q_w} + \m{v_{p}}\\
\m{y}=\m{h}(\m{x},\m{u})   + \m{{D}_w} \m{q_w},\qquad  
\end{subnumcases}
where the state, output, and input vectors are detailed as follows
\begin{align*}
\m x &= \bmat{\delta \quad \omega \quad e'_{\mathrm{q}} \quad e'_{\mathrm{d}}\quad V_{\mathrm{R}} \quad E_{\mathrm{fd}} \quad R_{\mathrm{f}}\quad tg_{\mathrm{1}}\quad tg_{\mathrm{2}}\quad tg_{\mathrm{3}}}^{\top} \\
\m y &= \bmat{e_{\mathrm{R}} \quad e_{\mathrm{I}}}^{\top}, \quad \m u = \bmat{i_{\mathrm{R}} \quad i_{\mathrm{I}} }^{\top}.
\end{align*}
The dynamic response is generated via a similar manner as in the \nth{4}-order model with the exception that here we consider Generator 14 as it possesses a bigger damping coefficient than Generator 4. This consequently gives a reduced oscillation for the dynamic response, especially after the fault. The simulations are performed before, during, and after the fault for 15 seconds.
For this model we simulate the controller dynamics and study how the observer tracks the steady state response of the generator.  
To compute the observer gain matrix, we again solve problem \eqref{eq:l_inf_theorem} as LMIs by setting $\nu_4 = 1$ and $\nu_2 = 30$. The performance matrix is set to be $\mZ = 2\times 10^{-4}\mI$ with matrix $\mC$ is constructed in a similar fashion as the one in Case 1. Since there is no closed-form expression for the \nth{10} order model, the Lipschitz constants $\gamma_f$ and $\gamma_h$ are chosen to be $200$ and $20$. The observer design is not sensitive to values for $\gamma_{f,h}$; different values yield similar performance.

\begin{figure}[t]	\vspace{-0.0cm}\hspace{0.0cm}\centerline{\includegraphics[keepaspectratio=true,scale=0.4]{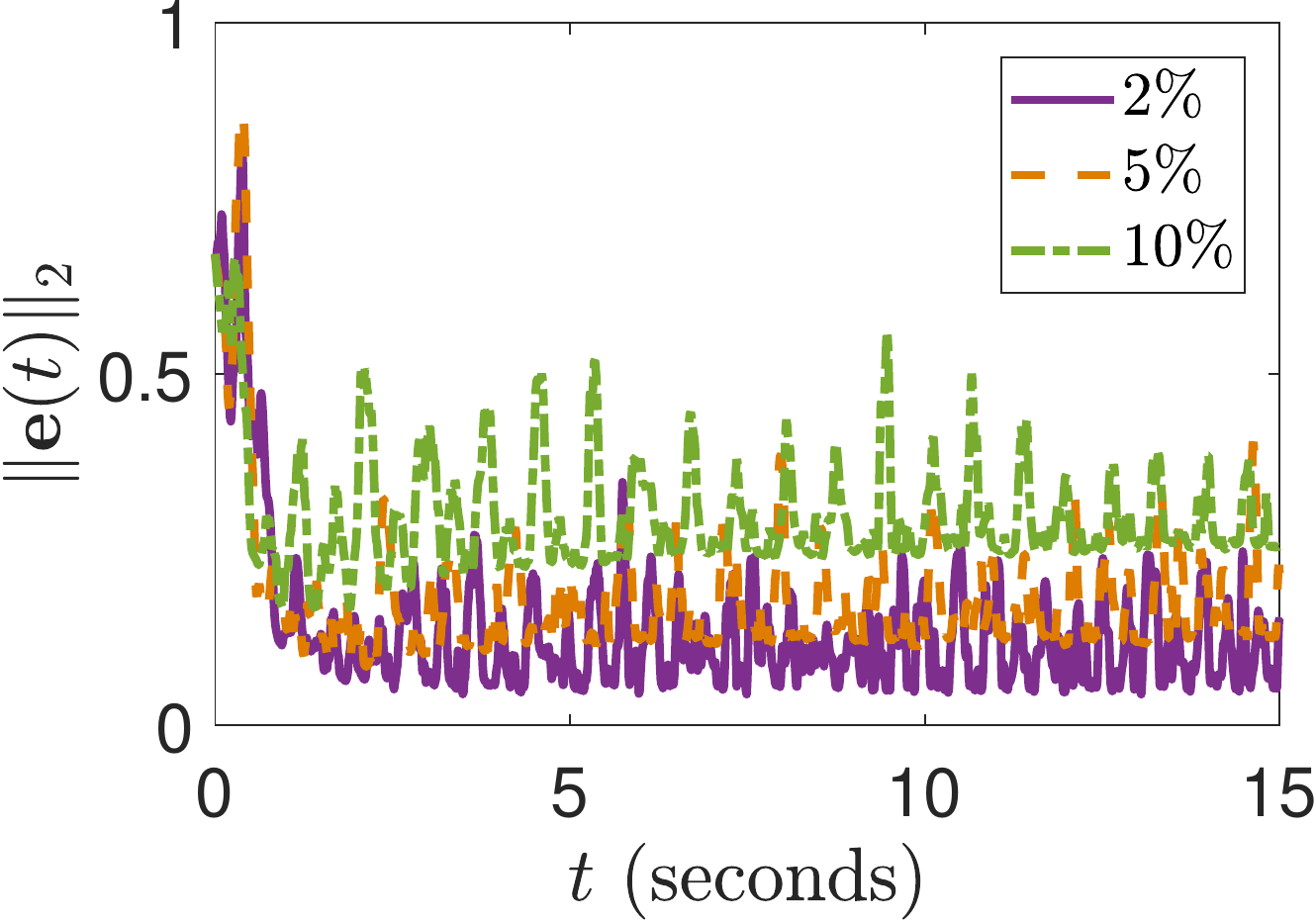}}
	\vspace{-0.7cm}
	\caption{Estimation error norm for Case 2 with $2$\%, $5$\%, and $10$\% model uncertainty.} 
	\label{fig:error_norm_mod_unc}
\end{figure}

\begin{figure}[t]	\vspace{-0.1cm}\hspace{0.0cm}\centerline{\includegraphics[keepaspectratio=true,scale=0.33]{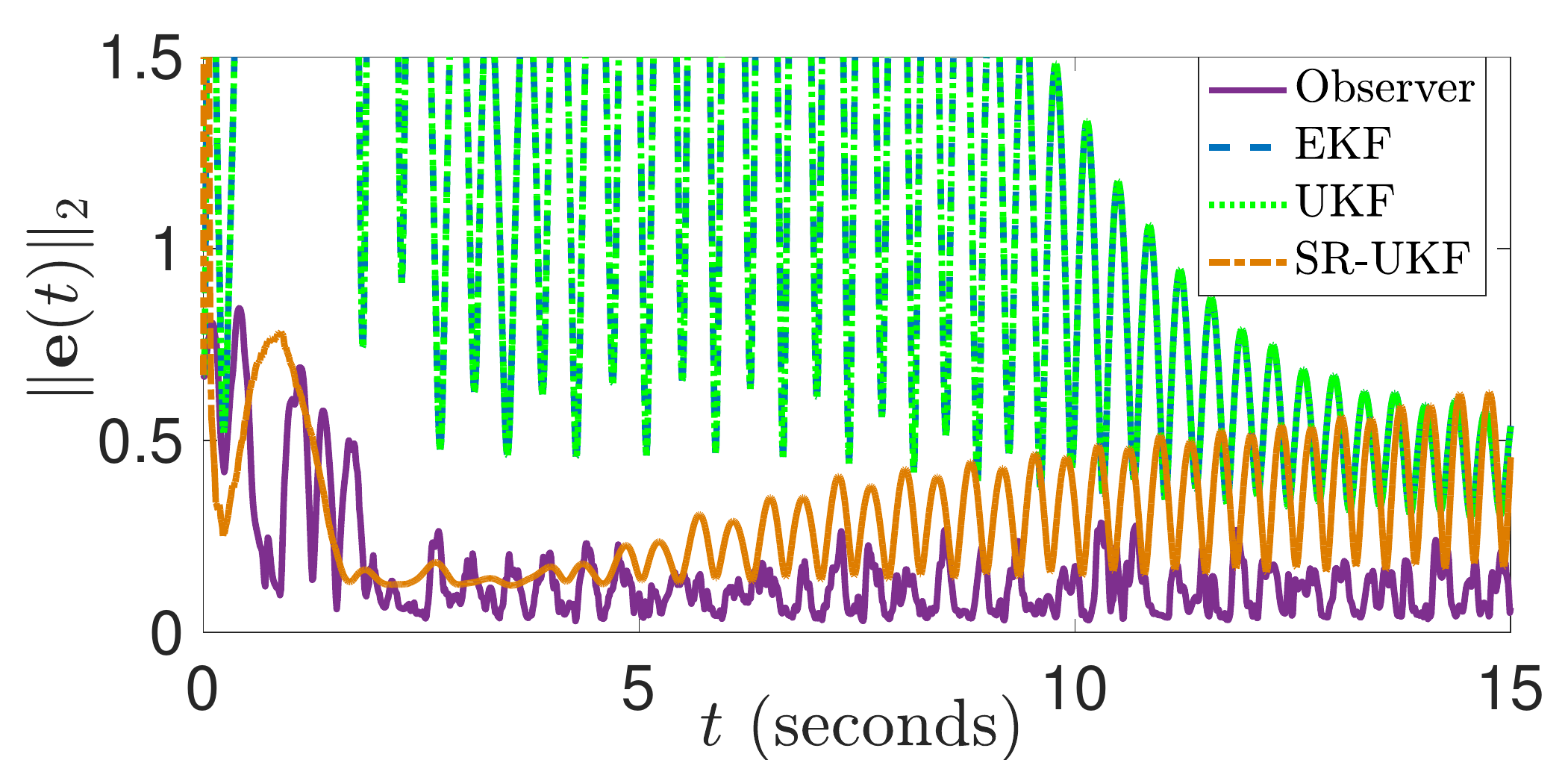}}
	\caption{The estimation error norm $\Vert\m e(t)\Vert_2$ for observer, EKF, UKF, and SR-UKF considering Case 2.}
	\label{fig:new_scen2_KF}
	\vspace{-0.2cm}
\end{figure}

\begin{figure}[t]
	\centering
	{\includegraphics[keepaspectratio=true,scale=0.35]{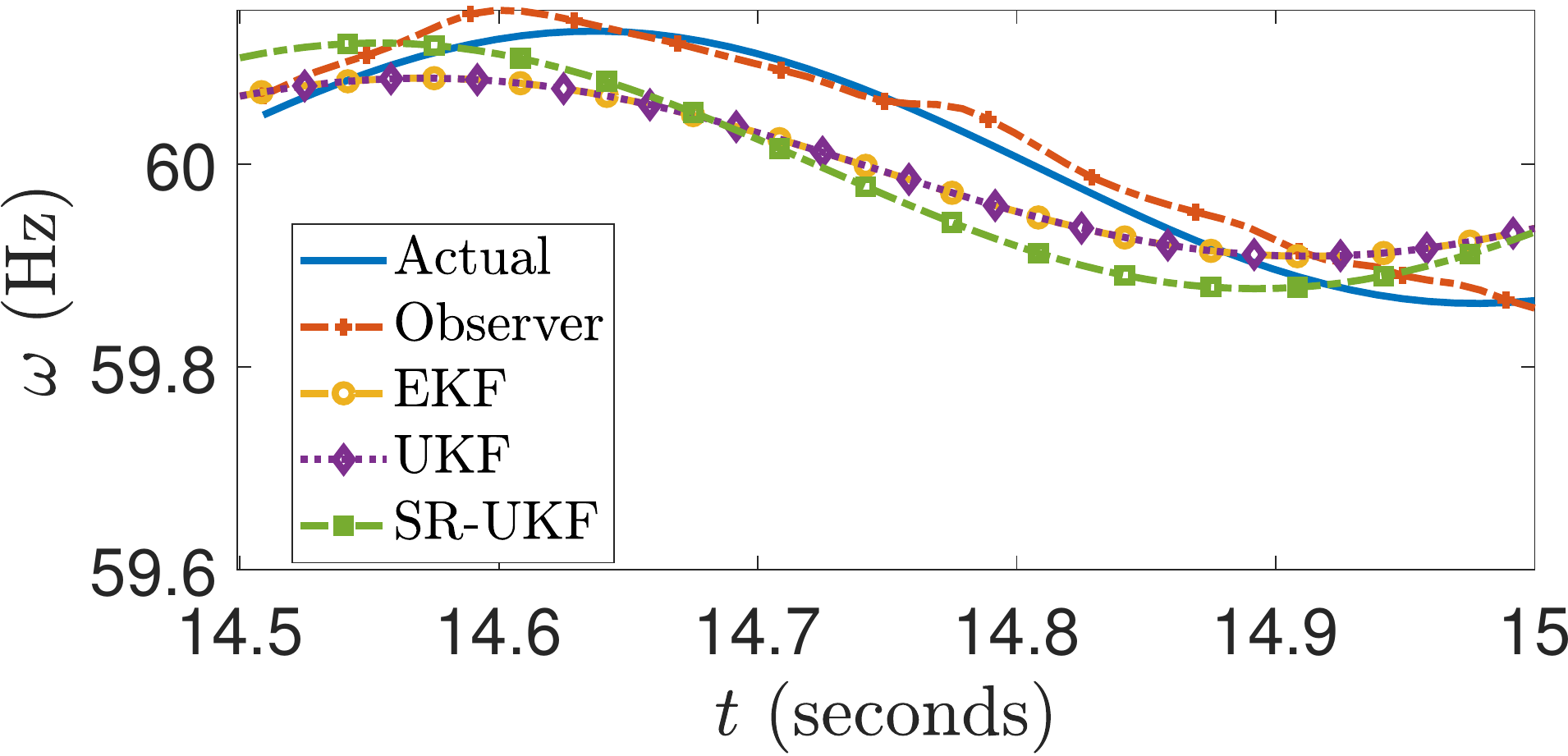}\label{fig:scen2_KF-1}}{\vspace*{-0.1cm}}
	\noindent \caption{Estimation results on generator's frequency for observer, EKF, UKF, and SR-UKF for Case 2 for the final half a second in the DSE.}\label{fig:scen2_KF}
\end{figure}

\begin{figure}[t]	\vspace{-0.0cm}\hspace{0.0cm}\centerline{\includegraphics[keepaspectratio=true,scale=0.35]{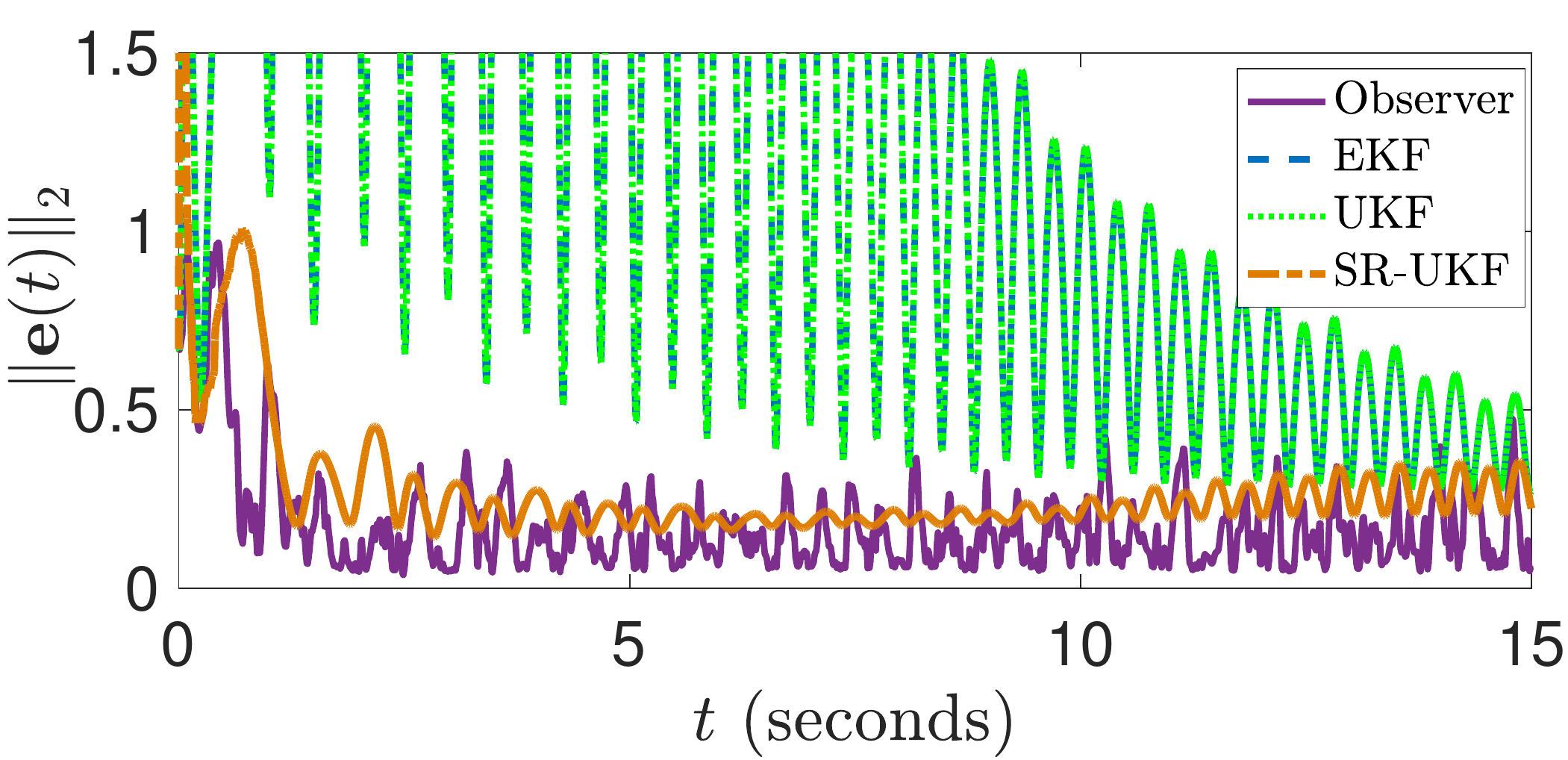}}
	\caption{The estimation error norm $\Vert\m e(t)\Vert_2$ for observer, EKF, UKF, and SR-UKF considering Case 3 with $2\%$ model uncertainty. }
	\label{fig:new_scen2_KF_laplace}
\end{figure}

\begin{figure}[t]	\vspace{0.0cm}\hspace{0.0cm}\centerline{\includegraphics[keepaspectratio=true,scale=0.40]{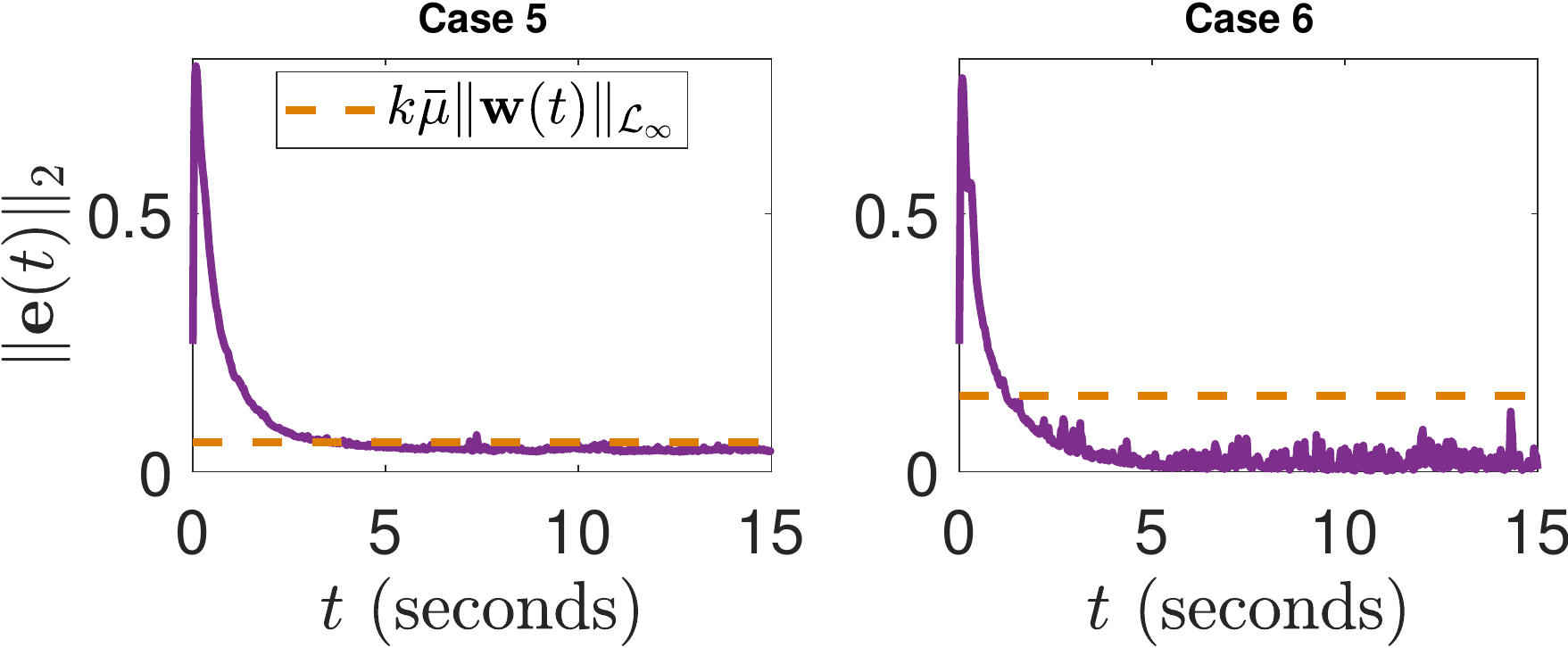}}
	\caption{The estimation error $\Vert\m e(t)\Vert_2$ and scaled worst-case disturbance $k\bar{\mu}\Vert\m w(t)\Vert_{\mathcal{L}_{\infty}}$ ($k = 5\times 10^{3}$) for Cases 5 and 6.}
	\label{fig:new_scen_z_all_10_ord}
\end{figure}

\begin{figure}[!t]
	\centering
	\subfloat[]{\includegraphics[keepaspectratio=true,scale=0.4]{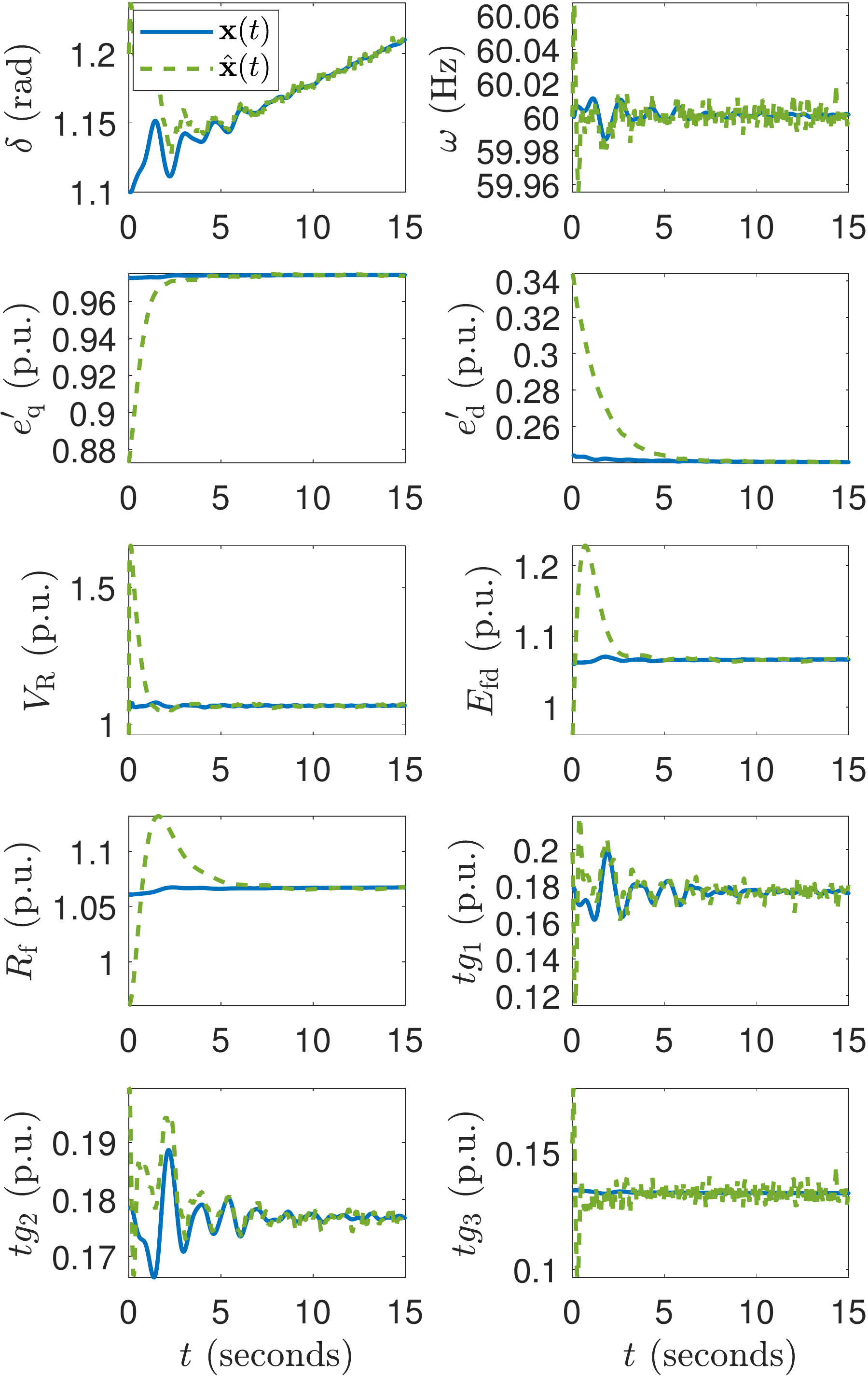}\label{fig:scen5-1}}{\vspace*{-0.15cm}}
	\subfloat[]{\includegraphics[keepaspectratio=true,scale=0.4]{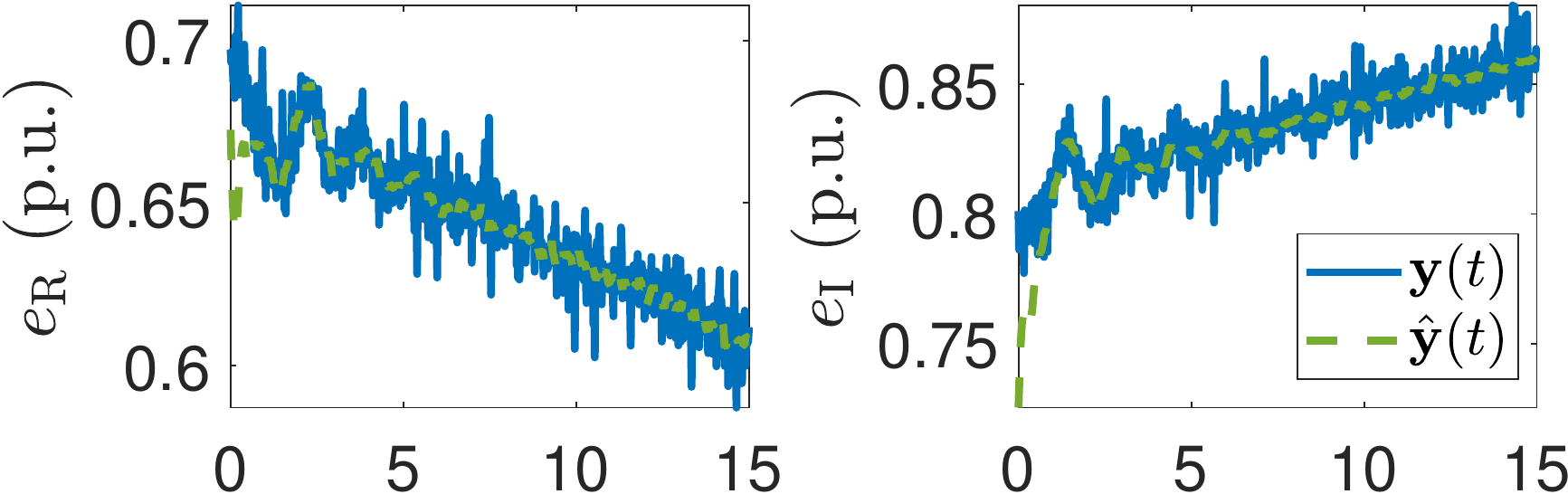}\label{fig:scen5-2}}{\vspace*{-0.1cm}}
	\noindent \caption{Actual and estimated states (a) and two outputs (b) for Case 6.}\label{fig:scen5}
\end{figure}

In this numerical test we consider the following two cases:

\noindent $\bullet$ \textit{Case 5:} This case considers Gaussian process and measurement noise in which the process noise covariance matrix is diagonal which entries are the square of $5$\% of the largest state changes and the measurement noise covariance matrix is also diagonal with variance $0.01^2$. The initial conditions of the observer are set to
\begin{align*}
 \hat{\m x}_{0}^{(1)} = [&0.9998\quad \omega_0 \quad 1.0729 \quad 0.1441\quad 1.1611 \quad \\ &1.1609\quad 1.1609\quad 0.1585\quad 0.1585\quad 0.1139]^{\top}.
\end{align*}

\noindent	$\bullet$ \textit{Case 6:} Similar to Case 5 but with Laplace measurement noise with $s = 0.01$ (see Section \ref{ssec:test1}) and a different estimator initial condition, which is
\begin{align*}
 \hat{\m x}_{0}^{(2)} = [&1.1998\quad \omega_0  \quad 0.8729\quad 0.3441\quad 0.9611 \quad  \\  &0.9609 \quad 0.9609\quad 0.1985\quad 0.1985\quad 0.1539]^{\top}.
\end{align*}
Note that both observer's initial condition in Case 5 and 6 is different than generator's initial condition for the generator:
\begin{align*}
\m x_{0} = [&1.0998\quad  \omega_0\quad 0.9729\quad 0.2441\quad 1.0611 \quad  \\  &1.0609 \quad 1.0609\quad 0.1785\quad 0.1785\quad 0.1339]^{\top}.
\end{align*}
Fig. \ref{fig:new_scen_z_all_10_ord} shows the estimation error norm for the two cases. The computed performance index for both cases are $\bar{\mu} = 5.9026\times 10^{-4}$, which acts as an upper bound on the optimal index, while the worst-case disturbance norm for Case 5 is $\norm {\m w(t)}_{\mathcal{L}_{\infty}} = 0.0191$ and for Case 6 is $\norm {\m w(t)}_{\mathcal{L}_{\infty}} = 0.0496$. 
The estimation results for the states and outputs are illustrated in Fig. \ref{fig:scen5} for Case 6---the results for Case 5 are similar thus not shown here for brevity. Despite of the presence of process, input, and measurement noise, the proposed observer is able to give adequate estimates with relatively small errors.

\vspace{-0.4cm}
\subsection{Discussion On the Derived Upper and Lower Bounds }

\begin{table}[t]
	\footnotesize	\centering \renewcommand{\arraystretch}{1.3}
\color{black}	\caption{Upper and Lower Bounds on the Performance Index}
	\begin{tabular}{l|l|l|l|l}
		\hline
		Model	& $\barbelow{J}$  & $\bar{J}$ & $\bar{J}-\barbelow{J}$& $\bar{\mu}$ \\ \hline \hline
		\nth{4}-order	& $4.5\times 10^{-8}$ & $1.7\times 10^{-7}$ & $1.2\times 10^{-7}$ &  $4.2\times 10^{-4}$\\ \hline
		\nth{10}-order	& $1.1\times 10^{-7}$ & $3.4\times 10^{-7}$ & $2.4\times 10^{-7}$ & $5.9\times 10^{-4}$ \\ \hline
	\end{tabular}\label{tab:upper_lower}
\end{table}

In this section we briefly investigate the applicability of the derived upper and lower bounds on the optimal solution for the $\mc{L}_{\infty}$ observer design problem. This relates to the discussion in Section~\ref{sec:observer-design}; see Theorems~\ref{l_inf_theorem} and~\ref{l_inf_relax}. Specifically, we compare the derived bounds $\bar{J}$ and $\barbelow{J}$ for  both \nth{4}-order and \nth{10}-order model. Table \ref{tab:upper_lower} shows the upper bounds, lower bounds, and gaps on the upper and lower bounds for the \nth{4}-order and \nth{10}-order model. 
Note that we clearly have $\barbelow{J} \leq J^* \leq \bar{J}$ with gap no bigger than $10^{-6}$, implying that the optimal performance $\mu^*$ is indeed very close to the computed---and utilized in the simulations---performance index $\bar{\mu}$. 
This is important in two ways. Firstly, the results corroborate the theoretical assertions made in Theorem~\ref{l_inf_relax} and secondly, they entail that the presented state estimation results are close to being optimal in the $\mc{L}_{\infty}$ stability sense as defined by Definition~\ref{L_inf_stability}. 
}

\section{Conclusion, Paper's Limitations, Future Work} \label{conclusion}
\color{black}
\vspace{-0.0cm}

This paper introduces a new observer design, developed using the notion of $\mc{L}_{\infty}$ stability, for DSE of uncertain synchronous generator models. The $\mc{L}_{\infty}$ observer provides a performance guarantee for the state estimation error norm relative to the worst case disturbance. As the observer gain remains constant, it can be computed offline and thus make it suitable for real time estimation. The numerical test results show that the performance of $\mc{L}_{\infty}$ observer is comparable (even better in some cases) to several mainstream DSE approaches considered in the literature, such as EKF, UKF, and SR-UKF.  

The paper's limitations are three-fold. First, this work only focuses on the estimation of generator's state and does not include the estimation of $T_{\mathrm{m}}(t)$ and $E_{\mathrm{fd}}(t)$.
Second, many other robust DSE methods have been proposed in the very recent literature. Some of these methods are based on derivative-free Kalman filtering~\cite{anagnostou2018derivative} and $\mc{H}_{\infty}$ UKF~\cite{Zhao2019tprs,Zhao2019TF}; see the survey paper~\cite{zhao2019power} and the discussions therein. Third, we only tackled the problem of decentralized state estimation for a single synchronous machine. To that end, future research directions will address the aforementioned limitations by \textit{(i)} deriving a robust observer for multi-machine network models, \textit{(ii)} theoretically accounting for unknown input estimation, and \textit{(iii)} performing comparative analysis of the many robust DSE methods in the recent, contemporary literature. 

Finally, designing the $\mc{L}_{\infty}$ observer considering a discrete-time machine and measurement model is an interesting future direction. This is needed seeing that the PMU measurements  are transmitted over digitized communication networks, and hence discrete-time observer is befitting for this application. This also guarantees a more consistent comparison with other KF-based estimation techniques that are simulated in discrete-time.

\vspace{-0.461cm}

\section*{Acknowledgments}
We gratefully acknowledge the constructive comments from the editor and the reviewers---they have all contributed positively to this work. We also acknowledge the financial support from the National Science Foundation through Grants 1728629 and 1917164, and  Cyber Florida under Collaborative Seed Award 3910-1009-00. \normalcolor

\vspace{-0.61cm}

\bibliographystyle{IEEEtran}	\bibliography{bibl}

\appendices

    \section{Proof of Theorem \ref{l_inf_theorem}}~\label{app:l_inf_theorem}
		Consider the estimation error dynamics \eqref{eq:est_error_dynamics} with performance output  $\m z(t) = \mZ \m e(t)$  and bounded disturbance $\m w(t)$. 
		Let $V(\m e) = \m e^{\top}\mP \m e$ be a Lyapunov function candidate  where $\mP \succ 0$. 
		From \cite[Theorem 1]{pancake2002analysis}, the estimation error dynamics \eqref{eq:est_error_dynamics} is $\mathcal{L}_{\infty}$ stable with performance level $\mu = \sqrt{\nu_1\nu_2+\nu_3}$ if there exist 
		constants $\nu_1,\nu_2,\nu_3\in \mathbb{R}_{+}$ such that 
		\begin{subequations}\label{eq:l_inf_lemma}
			\begin{align}
			\nu_1 \norm{\m w}_2^2 &< V(\m e) \;\Rightarrow \dot{V}(\m e) < 0 \label{eq:l_inf_lemma_1}\\
			\norm{\m z}_2^2 &\leq \nu_2V(\m e)+\nu_3\norm{\m w}_2^2,\label{eq:l_inf_lemma_2}
			\end{align}
		\end{subequations}
		for all $t\geq 0$. 
		Note that the first condition \eqref{eq:l_inf_lemma_1} is satisfied if there exists $\nu_4 > 0$ such that
		\begin{align*}
		\dot{V}(\m e) +\nu_4\left(V(\m e)-\nu_1\norm{\m w}_2^2\right)&\leq 0 \\
		\Leftrightarrow \dot{\m e}^{\top}\mP\m e + {\m e}^{\top}\mP\dot{\m e} + \nu_4 \m e^{\top}\mP\m e -\nu_4\nu_1\m w^{\top}\m w &\leq 0,
		\end{align*}
		which is equivalent to $\m \psi^{\top} \m\Theta \m\psi\leq 0$ where
		\begin{subequations}
			\begin{align*}
			\m \psi &= \bmat{\m e\\ \Delta\m f\\ \Delta \m{h_\mathrm{l}}\\ \m w^{\top}},
			\m\Theta = 
			\bmat{\m Q &*&*&*\\
				\mP&\mO&*&*\\
				-\mL^{\top}\mP&\mO&\mO&*\\
				\m {B_{w}}^{\top}\mP-\m {D_{{w}}}^{\top}\mL^{\top}\mP & \mO &\mO &-\nu_4\nu_1\mI}  \\
			\m Q &= \mA^{\top}\mP + \mP\mA -\mC^{\top}\mL^{\top}\mP-\mP\mL\mC +\nu_4\mP.	
			\end{align*}
		\end{subequations}
		Realize that $\m \psi^{\top} \m\Theta \m\psi\leq 0$ holds if $\m\Theta \preceq 0$ holds. Since the function $\m f(\cdot)$ is locally Lipschitz, then we have
		\begin{align*}
		\norm{\Delta\m f}_2^2 &\leq \gamma_f^2 \norm{\m e}_2^2 
		\;\Leftrightarrow \;\Delta\m f^{\top}\Delta\m f-\gamma_f^2\m e^{\top} \m e \leq 0, \nonumber 
		\end{align*}
		which is equivalent to $\m \psi^{\top} \m\Gamma \m\psi\leq 0$ where 
		\begin{align*}
	\m \Gamma = \mathrm{diag}\left(\bmat{-\gamma_f^2\mI&\mI&\mO&\mO}\right).
		\end{align*}
		Since $\m \psi^{\top} \m\Gamma \m\psi\leq 0$ for all admissible $\psi$, then $\m\Gamma \preceq 0$.  From the S-Lemma \cite{derinkuyu2006s}, then  $\m\Theta\preceq 0$ holds if there exists $\nu_5 \geq 0$ such that $\m\Theta - \nu_5\m\Gamma\preceq 0$. 
		Similarly, from the Lipschitz property of $\m {h_{\mathrm{l}}}(\cdot)$, then we have
		\begin{align*}
		\norm{\Delta\m {h_{\mathrm{l}}}}_2^2 &\leq \gamma_l^2 \norm{\m e}_2^2 
		\;\Leftrightarrow \;\Delta\m {h_{\mathrm{l}}}^{\top}\Delta\m {h_{\mathrm{l}}}-\gamma_l^2\m e^{\top} \m e \leq 0. \nonumber 
		\end{align*}
		This is equivalent to $\m \psi^{\top} \m\Pi \m\psi\leq 0$ where 
				\begin{align*}
		\m \Pi = \mathrm{diag}\left(\bmat{-\gamma_l^2\mI&\mO&\mI&\mO}\right).
		\end{align*}
		Since $\m \psi^{\top} \m\Pi \m\psi\leq 0$ for all admissible $\psi$, then $\m\Pi\preceq 0$. Again, by the S-Lemma, then  we have $\m\Theta - \nu_5\m\Gamma\preceq 0$ if there exists $\nu_6 \geq 0$ such that $\m\Theta - \nu_5\m\Gamma - \nu_6\m\Pi\preceq 0$, which is equivalent to \eqref{eq:l_inf_theorem_1} given that $\mY = \mP\mL$.
		Next, substituting $\m z(t) = \mZ \m e(t)$ to the second condition \eqref{eq:l_inf_lemma_2} yields
		\begin{align}
		\norm{\mZ \m e}_2^2 -\nu_2V(\m e)-\nu_3\norm{\m w}_2^2&\leq 0 \nonumber\\
		\Leftrightarrow\m e^{\top}\mZ^{\top}\mZ \m e-\nu_2\m e^{\top}\mP\m e -\nu_3 \m w^{\top} \m w &\leq 0 \nonumber \\
		\Leftrightarrow\bmat{\mZ^{\top}\mZ -\nu_2\mP & \mO \\ \mO &-\nu_3\mI} &\preceq 0. \nonumber
		\end{align}
		Using congruence transformation and applying  the Schur Complement to the above yields \eqref{eq:l_inf_theorem_2}. Thus, solvability of \eqref{eq:l_inf_theorem} ensures that the estimation error dynamics given in \eqref{eq:est_error_dynamics} is $\mathcal{L}_{\infty}$ stable with performance level $\mu = \sqrt{\nu_1\nu_2+\nu_3}$ and observer gain $\mL = \mP^{-1}\mY$. This completes the proof.
$\hfill\blacksquare$

\section{Proof of Theorem \ref{l_inf_relax}}\vspace*{-0.1cm}	~\label{app:theorem_SDP_relax}
	From \eqref{eq:l_inf_theorem}, first define $\m \Xi$, $\sigma$, and $\lambda$ as $\m \Xi = \nu_4 \m P$, $\sigma = \nu_4\nu_1$, and  $\lambda = \nu_1\nu_2$. Then, as $\m \Xi = \nu_4 \m P$ is equivalent to $\m \Xi_{i,j} = \nu_4 \m P_{i,j}$ for all $i,j$, defining
\begin{align*}
&\m E = \bmat{0& \frac{1}{2}& 0\\ \frac{1}{2}& 0&0\\0&0&0}, \; \m e = \bmat{0\\0\\ 1},\;\m \psi_{i,j} = \bmat{\nu_4 \\ \mP_{i,j}\\ \m\Xi_{i,j}}, \;\forall i,j,
\end{align*}
this relation can be written as
\begin{align}
\nu_4 \m P_{i,j}-\m \Xi_{i,j} &= \m \psi_{i,j}^{\top}\m E \m \psi_{i,j} - \m e^{\top}\m \psi_{i,j} \nonumber \\
&= \mathrm{trace}(\m E\m \Psi_{i,j}) - \m e^{\top}\m \psi_{i,j} = 0, \label{eq:proof_thm2_1} 
\end{align} 	
only when $\m \Psi_{i,j} = \m \psi_{i,j}\m \psi_{i,j}^{\top}$ holds. Realize that
\begin{align*}
\m \Psi_{i,j} = \m \psi_{i,j}\m \psi_{i,j}^{\top} \Leftrightarrow \m \Psi_{i,j} \succeq \m \psi_{i,j}\m \psi_{i,j}^{\top} \; \mathbf{and} \; \mathrm{rank}(\m \Psi_{i,j}) = 1,
\end{align*} 
then by using the Schur Complement such that 
\begin{align}
\m \Psi_{i,j} \succeq \m \psi_{i,j}\m \psi_{i,j}^{\top} \Leftrightarrow \bmat{\m\Psi_{i,j}&*\\\m \psi_{i,j}^{\top}&1} \succeq 0, \label{eq:proof_thm2_2}
\end{align} 
the bilinear term $\nu_4 \m P$ can be replaced with $\m \Xi$ while adding constraints \eqref{eq:proof_thm2_1}, \eqref{eq:proof_thm2_2}, and enforcing $\mathrm{rank}(\m \Psi_{i,j}) = 1$ for all $i,j$. 
Next, define new variables $\sigma$ and $\lambda$ such that $\nu_4\nu_1 = \sigma$ and $\nu_1\nu_2 = \lambda$. These equalities are equivalent to
\begin{subequations}\label{eq:proof_thm2_3_4} 
	\begin{align}
	\nu_4\nu_1 - \sigma &=\m \phi^{\top}\m E \m \phi -\m e^{\top}\m \phi = \mathrm{trace}(\m E \m \Phi)-\m e^{\top}\m \phi = 0 \label{eq:proof_thm2_3} \\
	\nu_1\nu_2 - \lambda &=\m \theta^{\top}\m E  \m \theta -\m e^{\top}\m \theta = \mathrm{trace}(\m E \m \Theta)-\m e^{\top}\m \theta = 0, \label{eq:proof_thm2_4} 
	\end{align}
\end{subequations}
where $\m \phi = \bmat{\nu_1 & \nu_4 & \sigma}^{\top}$ and $\m \theta = \bmat{\nu_1 & \nu_2 & \lambda}^{\top}$ only if we have $\m \Phi = \m \phi\m \phi^{\top}$ and $\m \Theta = \m \theta\m \theta^{\top}$. Note that 
\begin{align*}
\m \Phi = \m \phi\m \phi^{\top} \Leftrightarrow \m \Phi \succeq \m \phi\m \phi^{\top} \; \mathbf{and} \; \mathrm{rank}(\m \Phi) = 1 \\
\m \Theta = \m \theta\m \theta^{\top} \Leftrightarrow \m \Theta \succeq \m \theta\m \theta^{\top} \; \mathbf{and} \; \mathrm{rank}(\m \Theta) = 1
\end{align*} 
Again, from using Schur Complement such that
\begin{align}
\hspace{-0.3cm} \m \Phi \succeq \m \phi\m \phi^{\top} \Leftrightarrow \bmat{\m \Phi&*\\\m \phi^{\top}&1} \succeq 0,\;
\m \Theta \succeq \m \theta\m \theta^{\top} \Leftrightarrow \bmat{\m \Theta&*\\ \m \theta^{\top}&1} \succeq 0, \label{eq:proof_thm2_5}
\end{align} 
$\nu_4\nu_1$ and $\nu_1\nu_2$ can be replaced with $\sigma$ and $\lambda$ with the addition of constraints \eqref{eq:proof_thm2_3_4}, \eqref{eq:proof_thm2_5}, $\mathrm{rank}(\m \Phi) = 1$, and $\mathrm{rank}(\m \Theta) = 1$.  
By neglecting all of the nonconvex rank one constraints, we obtain an SDP relaxation as follows
\begin{subequations}\label{eq:l_inf_relax_p}
 		\begin{align}
&\minimize_{\substack{\m P, \m Y, \m\nu, \m \Xi, \m\Psi_{i,j}, \\ \m \Phi, \m \Theta, \lambda, \sigma}} \quad \lambda + \nu_3 \label{eq:l_inf_relax_p0}\\
&\subjectto \nonumber \\
&\hspace{-0.3cm}\bmat{ \mQ +\nu_5\gamma_f^2\mI+\nu_6\gamma_l^2\mI &*&*&*\\
	\mP & -\nu_5\mI&*&*\\
	-\mY^{\top} &\mO &-\nu_6\mI&*\\
	\m {B_{w}}^{\top}\mP-\m {D_{{w}}}^{\top}\mY^{\top}&\mO&\mO&-\sigma \mI} \preceq 0 \label{eq:l_inf_relax_p1}\\
&\hspace{-0.2cm} \eqref{eq:proof_thm2_1},\eqref{eq:proof_thm2_2},\eqref{eq:proof_thm2_3_4},\eqref{eq:proof_thm2_5}, \eqref{eq:l_inf_theorem_2}\label{eq:l_inf_relax_p5} \\
&\hspace{-0.2cm}\mP \succ 0, \;\m \Xi \succ 0, \;\nu_4 >0 ,\;\nu_{1,2,3,5,6} \geq 0, \;\lambda\geq 0,\sigma \geq 0,  \label{eq:l_inf_relax_p6}
\end{align}
\end{subequations}
where $\m Q = \mA^{\top}\mP + \mP\mA -\mC^{\top}\mY^{\top}-\mY\mC +\m \Xi$. The constraints in \eqref{eq:l_inf_relax_p} 
can be represented by 
\begin{subequations}\label{eq:l_inf_relax_pp}
\begin{align}
E(\m P, \m\nu, \m \Xi, \m\Psi_{i,j},\m \Phi, \m \Theta, \lambda, \sigma) &= 0 \label{eq:l_inf_relax_pp1}\\ L(\m P, \m Y, \m\nu, \m \Xi, \m\Psi_{i,j},\m \Phi, \m \Theta, \lambda, \sigma) &\preceq 0, \label{eq:l_inf_relax_pp2}
\end{align}
\end{subequations}
where \eqref{eq:l_inf_relax_pp1} and \eqref{eq:l_inf_relax_pp2} represents all matrix equality and inequality constraints respectively.
As \eqref{eq:l_inf_relax_p} is less constrained than \eqref{eq:l_inf_theorem} due to relaxation, then $\lambda + \nu_3 \leq \mu^{*2}$ where $\mu^* = \sqrt{\nu_1^*\nu_2^* + \nu_3^*}$ is obtained from solving \eqref{eq:l_inf_theorem}. This ends the proof.
$\hfill \blacksquare$

\vspace{-1.25cm}

\begin{IEEEbiography}
	[{\includegraphics[width=1in,height=1.25in,clip,keepaspectratio]{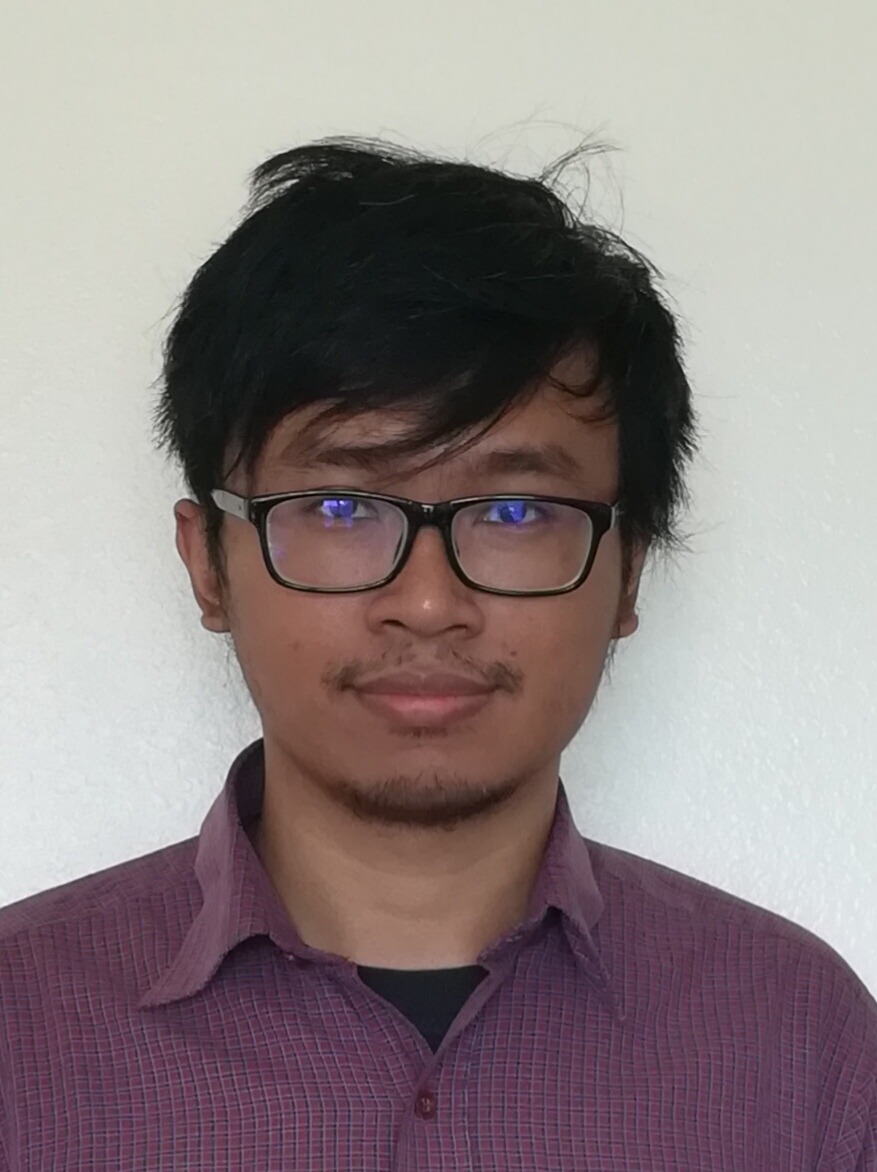}}]
	{Sebastian A. Nugroho} was born in Yogyakarta, Indonesia and received the B.S. and M.S. degrees in Electrical Engineering from Institut Teknologi Bandung (ITB), Indonesia in 2012 and 2014. He is currently a graduate research assistant and pursuing the Ph.D. degree in Electrical Engineering at the University of Texas, San Antonio (UTSA), USA. His main areas of research interest are control theory, state estimation, and engineering optimization with applications to cyber-physical systems.  
\end{IEEEbiography} 

\vspace{-1.5cm}
\begin{IEEEbiography}
	[{\includegraphics[width=1in,height=1.25in,clip,keepaspectratio]{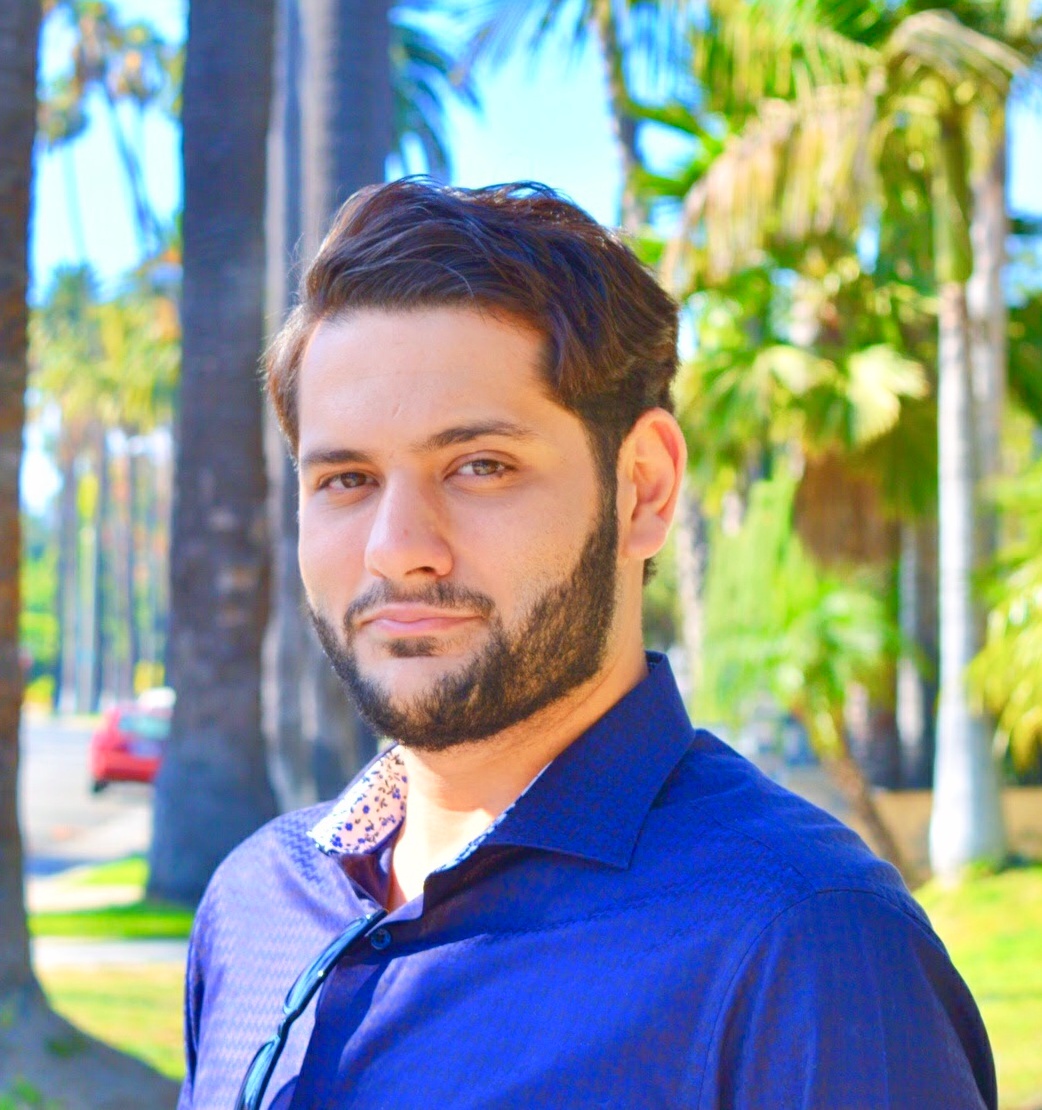}}]
	{Ahmad F. Taha} is  an assistant professor with the Department of Electrical and Computer Engineering at the University of Texas, San Antonio. He received the B.E. and Ph.D. degrees in Electrical and Computer Engineering from the American University of Beirut, Lebanon in 2011 and Purdue University, West Lafayette, Indiana in 2015. Dr. Taha is interested in understanding how complex cyber-physical systems (CPS) operate, behave, and \textit{misbehave}. His research focus includes optimization, control, and security of CPSs with applications to power, water, and transportation networks. Dr. Taha is an editor of IEEE Transactions on Smart Grid and the editor of the IEEE Control Systems Society Electronic Letter (E-Letter).
\end{IEEEbiography}

\vspace{-1.5cm}
\begin{IEEEbiography}
	[{\includegraphics[width=1in,height=1.25in,clip,keepaspectratio]{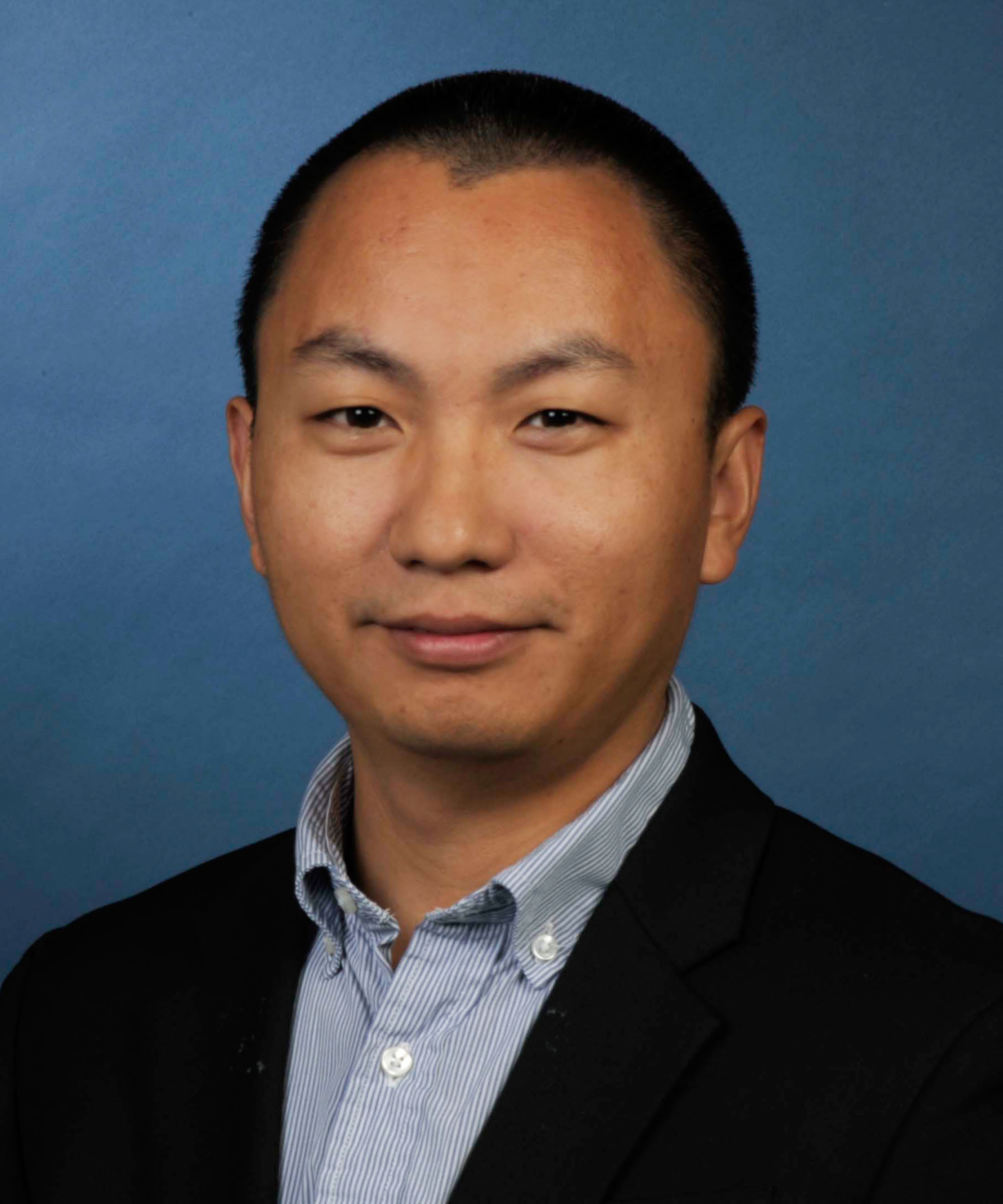}}]{Junjian Qi}  (S'12--M'13--SM'17)
	received the B.E. degree in electrical engineering, from Shandong University, Jinan, China, in 2008, and the Ph.D. degree in electrical engineering from Tsinghua University, Beijing, China, in 2013. He was a Visiting Scholar with Iowa State University, Ames, IA, USA, in 2012, a Research Associate with the Department of EECS, University of Tennessee, Knoxville, TN, USA, from 2013 to 2015, and a Postdoctoral Appointee with the Energy Systems Division, Argonne National Laboratory, Lemont, IL, USA, from 2015 to 2017. He is currently an Assistant Professor with the Department of Electrical and Computer Engineering, University of Central Florida, Orlando, FL, USA. Dr. Qi is the Secretary of the IEEE Working Group on Energy Internet and the IEEE Task Force on Voltage Control for Smart Grids. He is an Associate Editor of IEEE Access. His research interests include cascading blackouts, microgrid control, state estimation, synchrophasors, and cyber-physical system security.
\end{IEEEbiography}

\end{document}